\documentclass[fleqn,usenatbib,nofootinbib]{mnras}
\usepackage{graphicx}
\usepackage{amsmath,amssymb}
\usepackage{lastpage}
\usepackage[all]{hypcap}
\usepackage[T1]{fontenc}
\usepackage{float}
\usepackage{subfigure}
\usepackage{afterpage}
\usepackage[usenames]{color}
\usepackage{mathrsfs}
\usepackage{upgreek}
\usepackage{hyperref}
\usepackage{gensymb}
\usepackage{color}
\usepackage[dvipsnames]{xcolor}
\usepackage{longtable}
\usepackage{threeparttable}
\usepackage{multirow}
\usepackage{placeins}
\usepackage{bm}
\usepackage[capitalise]{cleveref}
\usepackage{fontawesome5}
\usepackage{enumitem}

\newcommand{\github}[1]{\href{#1}{\faGithubSquare}}
\newcommand{\esrgithub}{\github{https://github.com/DeaglanBartlett/ESR}}

\newcommand{\go}{g_\text{obs}}
\newcommand{\gb}{g_\text{bar}}
\newcommand{\Ug}{\Upsilon_\text{gas}}
\newcommand{\Ud}{\Upsilon_\text{disk}}
\newcommand{\Ub}{\Upsilon_\text{bulge}}
\newcommand{\dd}{\mathrm{d}}

\title[The functional form of the RAR]{On the functional form of the radial acceleration relation}
\author[Desmond, Bartlett \& Ferreira]{
Harry~Desmond$^{1}$\thanks{harry.desmond@port.ac.uk}, Deaglan J. Bartlett$^{2,3}$\thanks{deaglan.bartlett@physics.ox.ac.uk} and Pedro G. Ferreira$^3$\\
$^{1}$Institute of Cosmology \& Gravitation, University of Portsmouth, Dennis Sciama Building, Portsmouth, PO1 3FX, UK\\
$^{2}$CNRS \& Sorbonne Universit\'{e}, Institut d’Astrophysique de Paris (IAP), UMR 7095, 98 bis bd Arago, F-75014 Paris, France\\
$^{3}$Astrophysics, University of Oxford, Denys Wilkinson Building, Keble Road, Oxford OX1 3RH, UK\\
}

\pubyear{2022}

\begin{document}
\label{FirstPage}
\pagerange{\pageref{FirstPage}--\pageref{LastPage}}
\maketitle

\begin{abstract}
We apply a new method for learning equations from data---\emph{Exhaustive Symbolic Regression} (ESR)---to late-type galaxy dynamics as encapsulated in the radial acceleration relation (RAR). Relating the centripetal acceleration due to baryons, $\gb$, to the total dynamical acceleration, $\go$, the RAR has been claimed to manifest a new law of nature due to its regularity and tightness, in agreement with Modified Newtonian Dynamics (MOND). Fits to this relation have been restricted by prior expectations to particular functional forms, while ESR affords an exhaustive and nearly prior-free search through functional parameter space to identify the equations optimally trading accuracy with simplicity. Working with the SPARC data, we find the best functions typically satisfy $\go\propto\gb$ at high $\gb$, although the coefficient of proportionality is not clearly unity and the deep-MOND limit $\go \propto \sqrt{\gb}$ as $\gb \to 0$ is little evident at all. By generating mock data according to MOND with or without the external field effect, we find that symbolic regression would not be expected to identify the generating function or reconstruct successfully the asymptotic slopes. We conclude that the limited dynamical range and significant uncertainties of the SPARC RAR preclude a definitive statement of its functional form, and hence that this data alone can neither demonstrate nor rule out law-like gravitational behaviour.
\end{abstract}



\section{Introduction}
\label{sec:intro}

Kinematic measurements of galaxies relate their visible and dynamical masses, affording constraints on the distribution of dark matter and/or the behaviour of gravity. These measurements are simplest to perform for late-type galaxies supported predominantly by rotation, as the enclosed dynamical mass may be calculated from the centripetal acceleration and the law of gravity. Such studies have revealed a striking correlation between the enclosed baryonic and total dynamical mass assuming Newtonian gravity, dubbed the mass discrepancy--acceleration \citep{Sanders_1990, McGaugh_2004} or radial acceleration relation (RAR; \citealt{RAR}). It has been claimed that the RAR indicates that at high accelerations the Newtonian dynamical mass follows the baryonic mass (indicating little dark matter and the validity of Newtonian mechanics), while as acceleration drops below a new constant of nature $g_0 \approx 10^{-10}$ m s$^{-2}$ the dynamical mass increasingly exceeds the baryonic mass in a regular way.

One may attempt to understand these observations from either a dark matter or modified gravity perspective. In $\Lambda$CDM the difference between the dynamical and baryonic mass is due to the dark matter that makes up most of the mass of the galaxy. The RAR must therefore be explained by the relative distributions of dark and visible mass established by the process of galaxy formation. Interactionless cold dark matter is influenced only gravitationally by the baryonic mass so the emergence of the RAR must be somewhat fortuitous; it is not established directly by a baryon--dark matter coupling (although see \citealt{dipolar_dm, superfluid, dm-baryon_interaction} for alternative ideas). In contrast, the modified gravity (or modified inertia) interpretation posits a breakdown of Newtonian mechanics at low acceleration so that the dynamical mass inferred by a Newtonian analysis is not the true dynamical mass of the galaxy. The prototypical instantiation of this idea is Modified Newtonian Dynamics (MOND; \citealt{Milgrom_1, Milgrom_2, Milgrom_3}), in which the kinematic acceleration $\go$ follows the square root of the Newtonian acceleration $\gb$ in the weak-field regime. This enables the total dynamical mass of the galaxy to remain equal to the baryonic mass across galaxies' rotation curves, eliminating the need for dark matter in them. The MOND paradigm attempts to dispense with dark matter entirely, and has cosmologically viable relativistic extensions (most recently \citealt{skordis}). It is reviewed in \citet{Famaey_McGaugh} and \citet{Banik}.

Central to the dark matter--modified gravity debate in the context of galaxy dynamics is the functional form of the RAR. This is because MOND makes a very specific prediction (absent the external field effect: $g_\text{obs} = g_\text{bar}$ in the high-acceleration ``Newtonian regime'' and $g_\text{obs} \propto g_\text{bar}^{1/2}$ in the low-acceleration ``deep-MOND regime'') while dark matter could accommodate a range of possibilities depending on the effect of galaxy formation on halo density profiles, which remains highly uncertain \citep[e.g.][]{Duffy, Maccio, Grudic, Tenneti, Ludlow, Navarro, Keller}. The only potentially unambiguous prediction is that the RAR tends to $\go = \Omega_\text{m}/\Omega_\text{b} \: \gb$ at radii sufficiently large to encompass the cosmic baryon fraction, but it is unclear where or even if this occurs in galaxies. Thus, while the $\Lambda$CDM prediction for the full RAR can be tested only by applying potentially restrictive priors on galaxy formation effects \citep{DC_Lelli, Desmond_MDAR, Paranjape_Sheth}, a more direct route towards informing the dark matter--modified gravity debate is to test the MOND prediction, specifically the limiting behaviour at $g \ll g_0$ and $g \gg g_0$, the small intrinsic scatter and the lack of residual correlations.

Here we focus on the asymptotic behaviour. This can be assessed to some extent by fitting a functional form with free power-law slopes at both ends \citep{RAR}, but this assumes that the slope tends to a constant at each end and restricts to a specific part of the functional parameter space for which this is the case. These are in question when assessing the accuracy of the MOND prescription. A fully satisfactory fit should therefore make no such assumptions, eliminating potential confirmation bias and testing without any priors the assertion that the RAR implies no dynamically relevant dark matter at high $g$ and the deep-MOND limit at low $g$. We accomplish this here by means of a novel regression algorithm dubbed \emph{Exhaustive Symbolic Regression} (ESR; \citealt{ESR}), and hence assess the degree to which the RAR supports the tenets of MOND. Within the MOND paradigm, this method also enables optimisation of the ``interpolating function'' (IF) $\go = \mathcal{F}(\gb)$ between the two stipulated limits.

The structure of the paper is as follows. In Sec.~\ref{sec:data} we describe the RAR data that we use, and in Sec.~\ref{sec:method} our algorithm for generating functions and assessing their aptitude for describing the data. Sec.~\ref{sec:results} presents the results. In Sec.~\ref{sec:disc} we discuss the broader ramifications, potential remaining uncertainties and ways in which the programme could be furthered in the future. Sec.~\ref{sec:conc} concludes. Full details on ESR are given in the companion paper \citet{ESR}. Units not explicitly given are $10^{-10}$ m s$^{-2}$, and all logarithms are natural.


\section{Observational Data}
\label{sec:data}

We use the SPARC data set \citep{SPARC},\footnote{\url{http://astroweb.cwru.edu/SPARC/}} a compilation of 175 rotation curves from the literature combined with \textit{Spitzer} 3.6$\mu$m photometry. We apply the same quality cuts as the RAR study of \citet{RAR}, removing galaxies with quality flag 3 (indicating large asymmetries, non-circular motions and/or offsets between stellar and HI distributions) and those with inclinations $i<30\deg$, and points for which the quoted fractional uncertainty on the observed rotation velocity is greater than 10 per cent. This leaves 2,696 points from 147 galaxies.

\section{Method}
\label{sec:method}

We describe our method for generating and assessing trial functions in Sec.~\ref{sec:ESR}, and our likelihood function in Sec.~\ref{sec:loss}.
In Sec.~\ref{sec:mond} we outline our criteria for assessing whether a function displays MOND-like behaviour.

\subsection{Exhaustive Symbolic Regression}
\label{sec:ESR}

While algorithms for symbolic regression (SR)---the search for good functional descriptions of a dataset---are becoming mature, they remain fallible \citep{SR_review}. Unless the generating function of the data is known at the outset (in which case SR is not required), it is not possible to determine whether any SR algorithm has uncovered the best function. This motivated us to develop ``Exhaustive Symbolic Regression'' (ESR) which, given a set of basis functions, produces and evaluates \emph{every} possible function up to a given complexity of equation, defined here as the number of nodes in its tree representation. This enables a brute-force solution to relatively simple problems and provides a touchstone for assessing the results of stochastic algorithms at higher complexity. As shown in detail in \citet{ESR}, stochastic searches regularly fail to the best functions at even moderate complexity $\sim$6, so that we would not be confidant of obtaining the functional form of the RAR through any algorithm besides ESR.

Presented in full in the companion paper, ESR has two main steps: \textit{i)} generating, and optimising the parameters of, all functions up to a given complexity, and \textit{ii)} ranking these functions using an information-theoretic metric combining accuracy and simplicity. For part \textit{i}, the steps are:
\begin{enumerate}[label=(\arabic*),leftmargin=*]
\item Generate all possible trees containing a given number of nodes (equal to the complexity of functions considered).
\item Generate the complete set of such functions by 
decorating these trees with
all permutations of the operators from the operator list specified in advance,
utilising the constraints on the arity of the operator that can occupy a given node.
\item Simplify the functions and remove duplicates. Variants of the same function (e.g. $x(x+\theta_0)$ and $x^2 + \theta_0 x$) are however retained
as these may have different model complexities (used in step \textit{ii}, below). For each unique function the variant is retained that minimises this.
\item Determine the values of the free parameters appearing in the functions that maximise the likelihood of the data (see Sec.~\ref{sec:loss}).
\item Repeat for all complexities under consideration.
\end{enumerate}

The only degrees of freedom in this procedure are the maximum complexity considered (here set at 9 as higher complexity is computationally prohibitive) and the set of operators of which the functions are composed. Here we choose:\footnote{Many of these operators can manifestly be constructed from combinations of others. We include these repetitions to simplify the function trees, allowing a greater range of expressions up to the maximum complexity.}
\begin{itemize}
    \item \textbf{Nullary:} $\gb$, $\theta$
    \item \textbf{Unary:} $\exp$, sqrt, square, inv
    \item \textbf{Binary:} $+$, $-$, $*$, $/$, pow
\end{itemize}
where $\theta$ is a free parameter. We implicitly take the absolute value of the argument of any square root or power.

The result of this procedure is a list of all functions up to the maximum complexity (of which there are 2.24$\times10^7$), along with the parameter values that maximise the likelihood of the RAR data. As in regular regression, using the maximum likelihood as the model selection criterion would favour overfitting, whereby a function fits the data near-perfectly but generalises or extrapolates poorly. SR therefore typically uses a two-objective optimisation, where the second objective is the ``simplicity'' of the function. In the absence of a metric for trading accuracy (the first objective) with complexity, optimal functions form a ``Pareto front'' where accuracy cannot be increased without reducing simplicity and vice versa. Simplicity has been defined analogously to model complexity (the number of nodes in the tree representation; e.g. in \texttt{PySR};~\citealt{PySR}), among others, but such definitions are typically arbitrary and thus compromise the objectivity of the regression results.

To remedy this, part \textit{ii} of ESR implements the \emph{minimum description length principle} (MDL; \citealt{RISSANEN1978, MDL_review1, MDL_review2}) as a model selection criterion, which has an information-theoretic motivation and provides a natural framework for making commensurable the two objectives. MDL states that functions are preferred to the extent that they compress the data, i.e. minimise the number of bits required to communicate the data with the aid of the function. We implement this with a two-step code in which the description length (also called codelength) is comprised of a component describing the function and a component describing the residuals of the data around the function's expectation. We use the Shannon--Fano coding scheme for the latter \citep{cover_thomas}, and for the former include contributions both from the structure of the function (penalising those employing more operators) and from the free parameters (penalising more parameters, especially ones that must be specified to high precision to achieve a high likelihood). The overall codelength of the compressed data, $L(D)$, is derived in sec. 3 of \citet{ESR}:
\begin{equation}
    \label{eq:length_final}
    \begin{split}
        L(D) &= L(D|H) + L(H) \\
        &= -\log(\mathcal{\hat{L}}) + k\log(n) - \frac{p}{2} \log(3) \\
        & \quad + \sum_i^p \left( \frac{1}{2}\log(\hat{I}_{ii}) + \log(|\hat{\theta}_i|) \right) + \sum_j \log(c_j),
    \end{split}
\end{equation}
where $L$ is the description length, $D$ the dataset, $H$ the hypothesis (i.e. function in question), $\mathcal{L}$ the likelihood, $\theta$ a free parameter of the function, $k$ the number of nodes in the function's tree representation, $n$ the number of unique operators involved, $p$ the total number of free parameters, $I$ the Fisher information matrix of the parameters and $c_j$ any constant natural numbers generated by simplifications. A hat denotes evaluation at the maximum-likelihood point. With all logarithms natural, this is the number of nats required to communicate the data with the aid of the function. $L(D)$ supports a probabilistic interpretation over function space that generalises the likelihood: the relative probability of a function is $\exp(-L(D))$ \citep{MDL_review2}.

The structure of the function alone determines the $k\log(n)$ term, but the remaining terms require the free parameters to be numerically optimised to maximise the likelihood (which we use interchangeably with minimising the loss).\footnote{The $p\log(3)/2$ term is only affected by the numerical optimisation if any parameters are set to 0 due to their maximum-likelihood values being less than 1 precision unit from 0 (see sec. 3 of \citealt{ESR}).} We now describe our choice of likelihood for the RAR data.

\subsection{Loss function}
\label{sec:loss}

As is typical (e.g. \citealt{RAR}), we assume that $g_\text{bar}$, $g_\text{obs}$ and their uncertainties are uncorrelated across the dataset. We further assume that the true $\gb$ and $\go$ values, denoted $g_\text{bar}^\text{t}$ and $g_\text{obs}^\text{t}$, generate the observed values with lognormal probability distributions centred at the true values with widths given by their uncertainties $\delta \gb$ and $\delta \go$. Following \citet{RAR}, we fix the mass-to-light ratios $\Upsilon_\text{gas}=1.33$, $\Upsilon_\text{disk}=0.5$ and $\Upsilon_\text{bulge}=0.7$ and assign them 10, 25 and 25 per cent uncertainties respectively, summing these in quadrature to estimate $\delta V_\text{bar}$ and hence $\delta g_\text{bar}$ (assuming no uncertainty in radial position). We likewise assume the uncertainties on distance $D$ and inclination $i$ to be statistical and hence sum their contributions in quadrature with the quoted statistical uncertainty on $V_\text{obs}$ according to \citet[eq. 2]{RAR} to estimate $\delta g_\text{bar}$.

The likelihood of 
an observation given the function in question, $f(\gb^\text{t})$, is then:
\begin{equation}
    \label{eq:loss}
    \begin{split}
        &\mathcal{L}(\log(\go)) = \int^\infty_{-\infty} \mathcal{L}(\log(\go)|\log(\gb^\text{t})) \: \mathcal{L}(\log(\gb^\text{t})) \: \dd \log(\gb^\text{t}) \\
        &= \frac{1}{2 \pi \: \delta\log(\gb) \: \delta\log(\go)} \int^\infty_{-\infty} \exp \left(-\frac{\left(\log(\go)-\log(f(\gb^\text{t}))\right)^2}{2 \: \delta\log(\go)^2} \right) \\
        &\times \exp \left(-\frac{\left(\log(\gb^\text{t})-\log(\gb)\right)^2}{2 \: \delta\log(\gb)^2} \right) \dd \log(\gb^\text{t})\\
        &\approx \frac{1}{\sqrt{2 \pi \sigma_\text{tot}^2}} \exp \left(-\frac{\left(\log(\go)-\log(f(\gb))\right)^2}{2 \sigma_\text{tot}^2} \right),
    \end{split}
\end{equation}
where
\begin{equation}
\label{eq:sigmatot}
    \sigma_\text{tot}^2 \equiv \delta\log(\go)^2 + \left(\frac{\dd\log(f(\gb))}{\dd\log(\gb)} \right)^2 \: \delta\log(\gb)^2.
\end{equation}
This is derived by keeping the leading-order term in the Taylor expansion of
$\log(f(\gb^\text{t}))$ around $\log(f(\gb))$
and therefore assumes this to be small relative to the rate of change of $f$. An advantage of working in $\log(\gb)-\log(\go)$ space as opposed to $\gb-\go$ is that, the RAR being roughly a product of power-laws, this minimises the error due to the first-order approximation. We find this to be good for all of the best functions.
The likelihood is then the product over all data points. We discuss in Sec.~\ref{sec:app_loss} the limitations of this likelihood model and how it could be improved.

\subsection{Assessing MOND}
\label{sec:mond}

\begin{figure}
  \centering
  \includegraphics[width=0.47\textwidth]{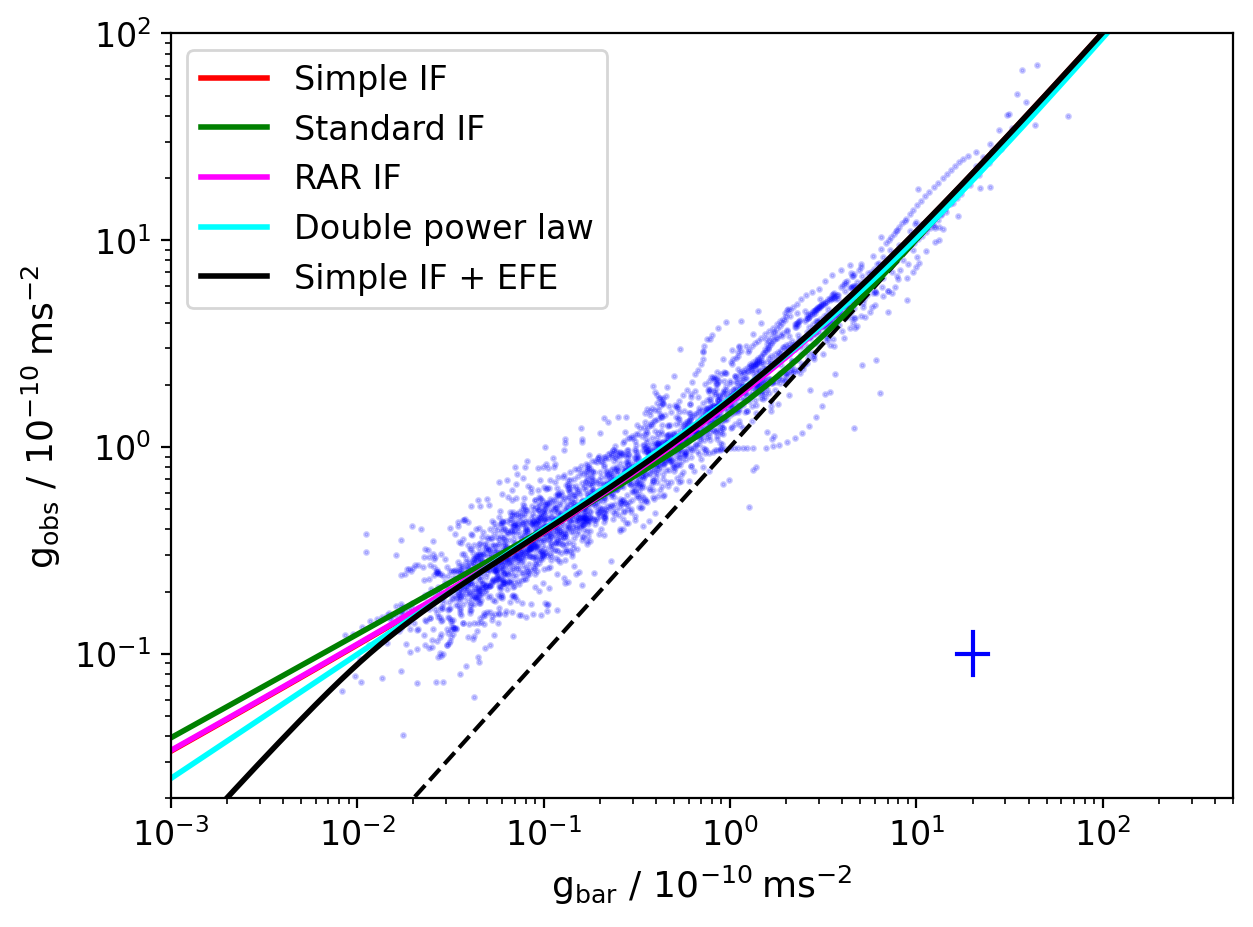}
  \caption{The Simple, Standard and RAR IFs, a double power law, and Simple IF with a global external field strength in AQUAL, overlaid on the SPARC data (blue points). The parameters are set to their maximum-likelihood values shown in Table~\ref{tab:real_results}. The dashed black line shows the one-to-one relation (Newtonian limit) and the cross in the lower right shows the average uncertainty size.
  }
  \label{fig:IFs}
\end{figure}

The core of the MOND paradigm is that $\go = \gb$ at $g \gg g_0$ and $\go = \sqrt{\gb g_0}$ at $g \ll g_0$ \citep{Milgrom_1, Milgrom_2, Milgrom_3}. This implies 
\begin{align}\label{eq:limits}
s \equiv \frac{\text{d}\log(\go)}{\text{d}\log(\gb)} =
  \begin{cases}
   1,   &\quad \gb \rightarrow \infty\\
   1/2, &\quad \gb \rightarrow 0.
  \end{cases}
\end{align}
Common choices for the IF covering the intermediate region $g \approx g_0$ include
\begin{itemize}
    \item \textbf{``Simple'':} $\go = \gb/2 + \sqrt{\gb^2/4 + \gb g_0}$ \citep{simple},
    \item \textbf{``Standard'':} $\go = \frac{1}{\sqrt{2}} \sqrt{\gb^2 + \sqrt{\gb^2 (\gb^2 + 4g_0^2}})$ \citep{Milgrom_2},
    \item \textbf{``RAR'':} $\go = \gb/(1-\exp(-\sqrt{\gb/g_0}))$ \citep{RAR}.
\end{itemize}
Fig.~\ref{fig:IFs} plots these functions on top of the RAR data for the best-fit values on the SPARC data (shown in the lower rows of Table~\ref{tab:real_results} in Sec.~\ref{sec:real_data}).
The Simple and RAR IFs are distinguished from the Standard IF principally by a more gradual transition between the Newtonian and deep-MOND regimes, although the Standard IF also prefers a significantly higher value of $g_0$.
While the basic MOND framework is not committed to any particular IF, it is committed to Eq.~\ref{eq:limits} providing an optimal description of the data. Our assessment of the theory will therefore be based on the extent to which the best functions (those with lowest description length) conform to these limits: any function that does so, in addition to possessing a coefficient of proportionality of unity in $\go\propto\gb$ at high $\gb$, may be considered a new MOND IF. (The low-$\gb$ coefficient of proportionality is $\sqrt{g_0}$ but $g_0$ is unknown a priori, so this does not supply an additional requirement.) Following \citet{RAR}, we will also consider a double power law fit:
\begin{equation}\label{eq:double_power_law}
\go = \theta_1 \: \left(1+\frac{\gb}{\theta_0}\right)^{\theta_2-\theta_3} \: \left(\frac{\gb}{\theta_0}\right)^{\theta_3}
\end{equation}
which has limiting logarithmic slopes of $\theta_3$ and $\theta_2$, and plot the best-fit in \cref{fig:IFs}. We define $s_- \equiv \lim_{\gb\rightarrow0} s$ and $s_+ \equiv \lim_{\gb\rightarrow\infty} s$.

The MOND interpretation of the RAR is complicated by the possibility of the external field effect (EFE), a breakdown of the strong equivalence principle due to the non-linear, acceleration-based modification to Newtonian mechanics \citep{Milgrom_1}. The EFE implies that otherwise identical galaxies in different external gravitational fields have different dynamics, which is a function of the external field strength $g_\text{ex}$ relative to $g_0$ and the internal field $g_\text{in}$. In the quasi-Newtonian regime $g_\text{in} < g_\text{ex} < g_0$, Kepler's laws are recovered with dynamical masses scaled by $g_\text{ex}/g_0$, while in the external field-dominated regime $g_\text{in} < g_0 < g_\text{ex}$, Newtonian mechanics are fully recovered \citep{Famaey_McGaugh}. This steepens the RAR at low $\gb$.

The precise effect of the EFE is difficult to calculate in general because it depends both on the underlying MOND theory and on a galaxy's morphology and orientation with respect to the external field direction. The most sophisticated fitting functions to MOND simulations are currently to be found in \citet{EFE_QUMOND} for QUMOND \citep{qumond} and \citet{Chae_Milgrom} for AQUAL \citep{aqual}. \citet{Paper_III} tested these expectations by fitting the SPARC rotation curves for the average external field strength, finding this to be in good agreement with independent estimates based on the baryonic mass surrounding the SPARC galaxies \citep{Paper_II} for AQUAL, but less so for QUMOND. We will therefore use AQUAL to explore the effect of the EFE on the expected low-$\gb$ slope and functional form more generally. \citet{Chae_Milgrom} eq. 15 gives

\small
\begin{eqnarray}\label{eq:efe}
&&\go = \gb \left(\frac{1}{2} + \left(\frac{1}{4} + \left(\left(\frac{\gb}{g_0}\right)^2 + (1.1 e_\text{N})^2\right)^{-\frac{1}{2}}\right)^\frac{1}{2}\right) \times \nonumber \\
&& \Biggl(1 + \tanh\left(\frac{1.1 e_\text{N}}{\gb/g_0}\right)^{1.2} \times \left(-\frac{1}{3}\right) \times \\
&& \frac{\left(\left(\left(\frac{\gb}{g_0}\right)^2 + (1.1 e_\text{N})^2\right)^{-\frac{1}{2}}\right) \left(\frac{1}{4} + \left(\left(\frac{\gb}{g_0}\right)^2 + (1.1 e_\text{N})^2\right)^{-\frac{1}{2}}\right)^{-\frac{1}{2}}}{1 + \left(\frac{1}{2} + 2\left(\left(\frac{\gb}{g_0}\right)^2 + (1.1 e_\text{N})^2\right)^{-\frac{1}{2}}\right)^\frac{1}{2}}\Biggr). \nonumber
\end{eqnarray}
\normalsize
This allows for variable disk thickness and scale length, and is azimuthally averaged to reduce sensitivity to the orientation of the field relative to the disk axis. It recovers the Simple IF as $e_\text{N}\equiv g_\text{ex}/g_0\rightarrow0$ and hence we refer to it as ``Simple IF + EFE''.

Note that, while some form of the EFE is generically predicted by MOND, in modified inertia formulations it may be very different (e.g. a function of the entire past trajectory of an object) or negligible~\citep{Milgrom_2011}. While there is evidence for the EFE in many systems~\citep{McGaugh_Milgrom,Haghi,Paper_I}, in others it appears conspicuously absent~\citep{Hernandez,Freundlich}. The black curve in Fig.~\ref{fig:IFs} shows the best fit to the data using Eq.~\ref{eq:efe}.

\subsection{Mock data generation}
\label{sec:mock_method}

To shed light on the significance of our results we apply ESR also to two stacks of mock data sets. We generate each mock data set using exactly the same number of points as the SPARC data, and with identical $\log{\gb}$, $\delta \log{\gb}$ and $\delta \log{\go}$ values, but with $\log{\go}$ generated using a MONDian function. This assumes $\gb^\text{t}$ equal to the SPARC $\gb$ (the maximum a priori estimate), $\log\gb$ for each mock realisation drawn from $\mathcal{N}(\log\gb^\text{t}, \delta \log\gb)$, and $\log{\go}$ from $\mathcal{N}(\mathcal{F}(\log\gb^\text{t}), \delta \log\go)$. To reduce the impact of noise in the mock data we apply ESR to a stack of 10 independent realisations.\footnote{The reason not to do more is that the time required for parameter optimisation scales with the number of data points, and this is already expensive at complexity 9. We assess convergence by fitting the complexity 4-6 equations to an independent stack of 10 realisations. We find that the order of equations when sorted by $L(D)$ is identical between the two stacks and that the difference between the description lengths of particular equations falls with complexity, from $\sim40$ at complexity 4 to $\sim5$ at complexity 6. This implies that the best equations, mostly at complexity 8 and 9, are not sensitive to the random number generation.} The only terms in the description length that depend on the dataset size are $\log(\mathcal{L})$ and the Fisher matrix $I$, both of which scale linearly. To make the results compatible with the real data we therefore divide these terms by 10.

The two mock data set stacks differ in the generating function $\mathcal{F}$. For the first, we use the RAR IF with the best-fit value on the data $g_0 = 1.127$ (see Sec.~\ref{sec:results}). This function is already known to describe the RAR well \citep{RAR} and has low enough complexity to be included (as a special case of a more general function, see below) in our function list. Since Eq.~\ref{eq:limits} is satisfied by construction in this case, evaluating it on the best functions from ESR will address the question of whether the dynamic range of the data is sufficiently high---and the uncertainties sufficiently low---to pick out unambiguously a correctly MONDian solution, as only in this case could one expect to obtain such behaviour for the real data were it generated by MOND.

The second stack is created using Eq.~\ref{eq:efe}. We adopt $g_0=1.2$ and $\langle g_\text{ex}\rangle=1.2\times10^{-2}$ ($e_\text{N}=0.01$), corresponding roughly to maximal clustering of unobserved baryons (as expected in a MOND cosmology and maximising agreement with the rotation curve fits;~\citealt{Paper_II}) and hence providing an upper bound on the impact of the EFE. This is similar to the value inferred in~\citet{Kyu_distinguishing} and~\citet{Paper_III} from fits to the SPARC rotation curves, and from our fit to the SPARC data in Table~\ref{tab:real_results}.\footnote{\citet{Chae_Milgrom} argue that Eq.~\ref{eq:efe} is only reliable with the inner points of galaxies' rotation curves removed. As we are interested only in the approximate effect of the EFE on the low-$\gb$ slope of the RAR, mainly sourced by outer rotation curve points, we do not apply a cut.
}

\section{Results}
\label{sec:results}


\subsection{SPARC data}
\label{sec:real_data}

\begin{figure}
  \centering
  \includegraphics[width=0.47\textwidth]{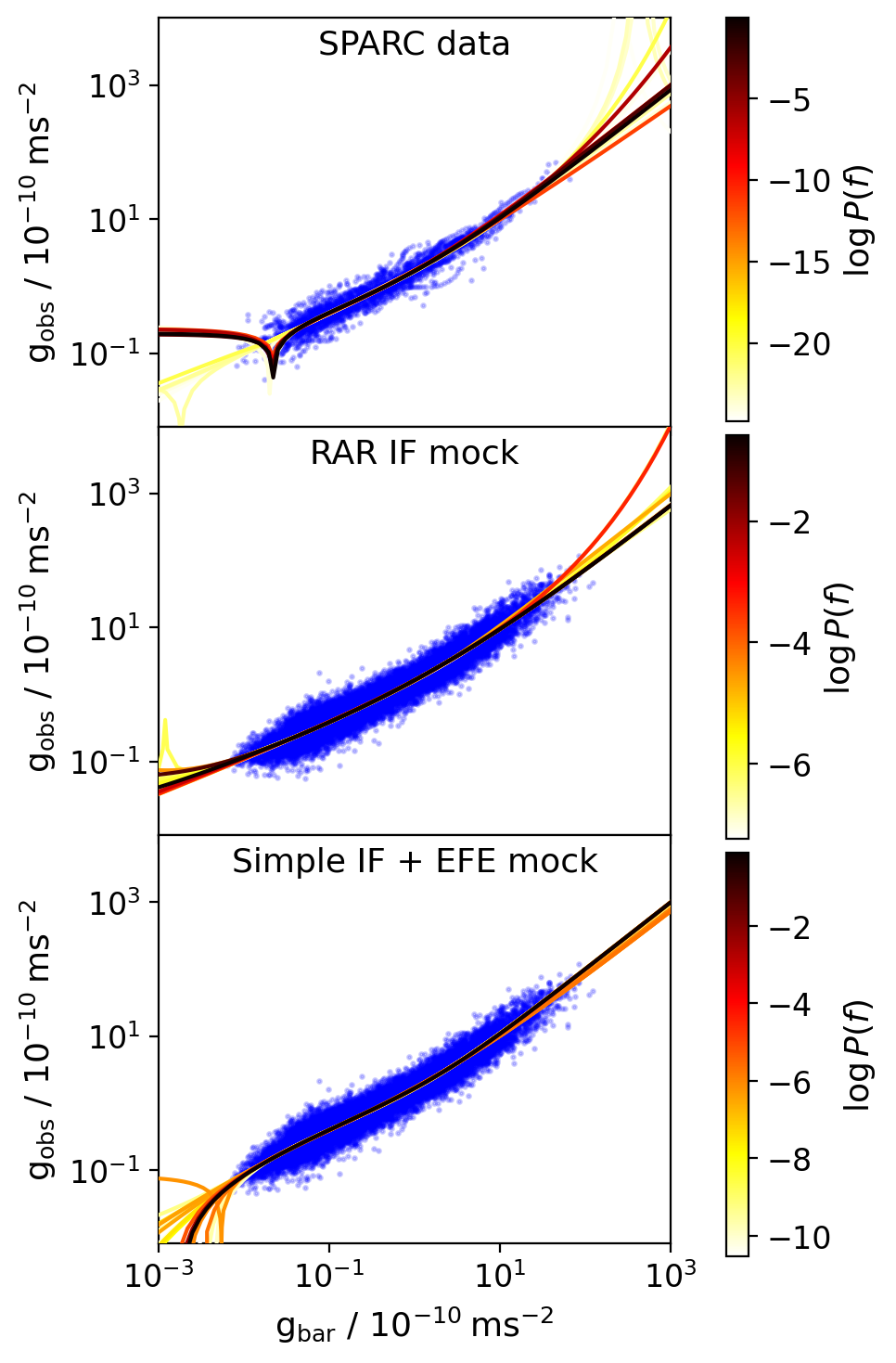}
  \caption{The top 20 functions found by ESR overlaid on the SPARC data (blue points), colour-coded by their relative probability in the full function list. The top panel fits the SPARC data, the middle panel mock data generated by the RAR IF, and the bottom panel mock data generated by the Simple IF with universal external field strength $g_\text{ex}=1.2\times10^{-12}$ m s$^{-2}$. The mock datasets are $10$ times larger than SPARC, although this is factored out in the description length calculation.
  }
  \label{fig:ESR_results}
\end{figure}

\begin{figure*}
  \centering
  \includegraphics[width=1.\textwidth]{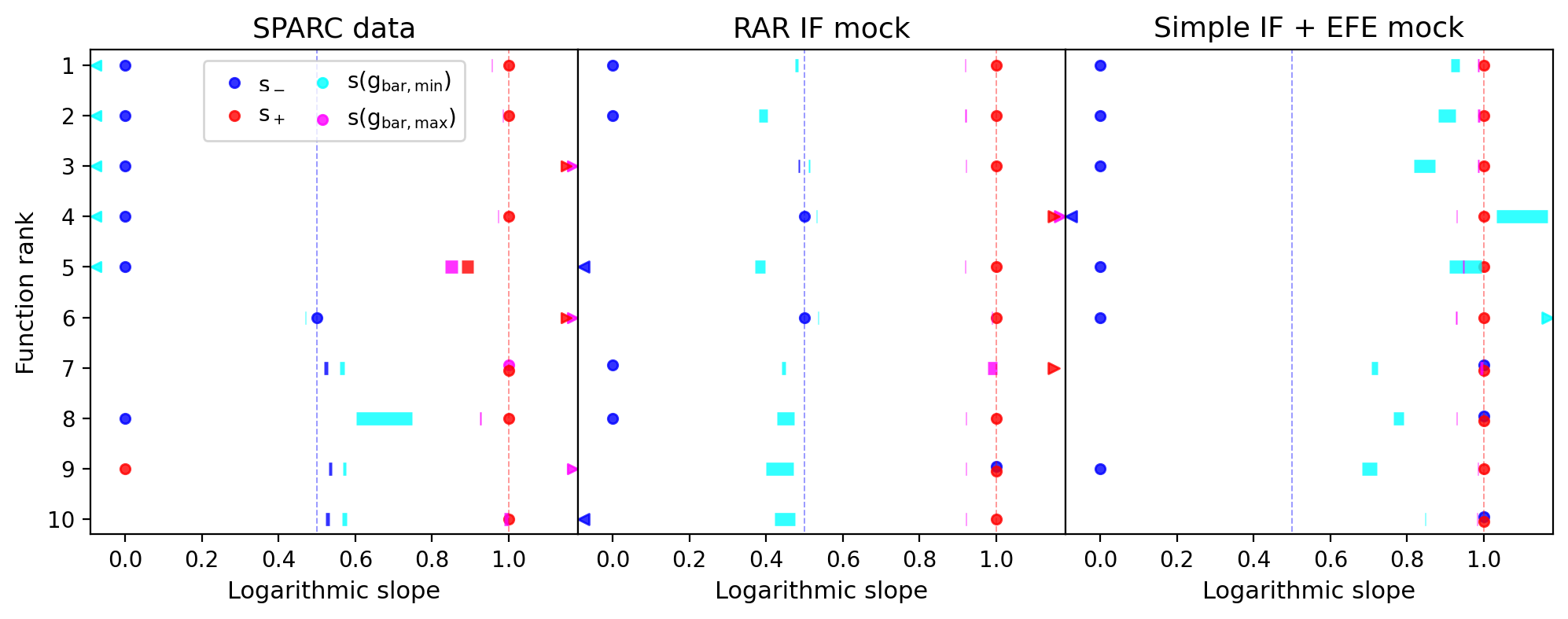}
  \caption{The logarithmic slopes $s \equiv \frac{d\log(\go)}{d\log(\gb)}$ of the top 10 ESR functions on each dataset, for comparison with the low- and high-$\gb$ MONDian expectations 1/2 and 1 respectively (blue and red vertical dashed lines). The blue and red points are the limiting slopes $s_- \equiv \lim_{\gb\to0^+} s$ and $s_+ \equiv \lim_{\gb\to\infty} s$, while cyan and magenta indicate the slopes at the minimum and maximum $\gb$ of the SPARC data (0.0083 and 65.4). In case a slope depends on a parameter value we show the 95\% confidence interval as a bar (often very thin), obtained from an MCMC fit. Arrowheads indicate points or bars beyond the range of the plot.}
  \label{fig:slopes}
\end{figure*}

\begin{figure*}
  \centering
  \includegraphics[width=1.\textwidth]{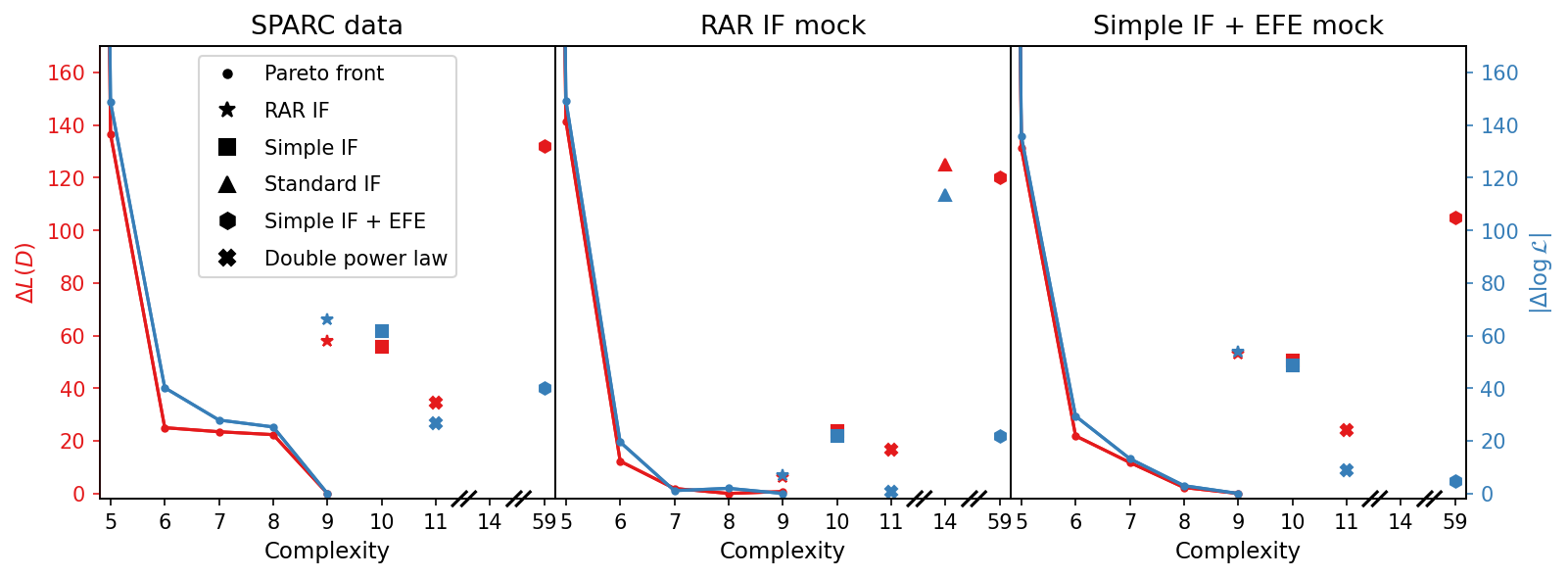}
  \caption{The Pareto fronts identified by ESR for the SPARC, RAR IF mock and Simple IF + EFE mock datasets, for both $\log(\mathcal{L})$ (blue) and total description length $L(D)$ (red). The quantities plotted have the minimum values subtracted so that the best results appear at 0. Also shown are the results of the RAR, Simple and Standard IFs, Simple IF + EFE and double power law fits. ESR significantly outperforms these ``by eye'' guesses, even for mock data generated from them. Short diagonal lines on the $x$-axis indicate breaks. In the left and right panels both red and blue points for the Standard IF at complexity 14 lie above the top of the plot.
  }
  \label{fig:Pareto}
\end{figure*}

We show in Table~\ref{tab:real_results} the statistics of the best functions found by ESR on the SPARC data. We split the codelength of Eq.~\ref{eq:length_final} into terms describing the residuals of the data around the functional expectation, the functional form and the parameter values as shown in the table footnotes. Below the horizontal line we give the results of the three MOND IFs, for which the free parameter corresponds to $g_0$, the double power law (Eq.~\ref{eq:double_power_law}) and the Simple IF + EFE (Eq.~\ref{eq:efe}). $P(f) \equiv \exp(-L(D))/\sum(\exp(-L(D)))$ is the probability of the function given its description length, where the sum is over all functions up to complexity 9. (Note that these values would be changed by low-$L(D)$ functions at higher complexity.) For reference, $L(D)$ for the raw data (corresponding to the hypothesis $\log\go=0$) is 53471, showing that significant compression is possible.

We find the best-fit $g_0$ value for the RAR IF to be 1.13, somewhat lower than the 1.20 quoted by \citet{RAR} although the data are the same. This is because \citet{RAR} used \texttt{scipy.odr} to perform the optimisation rather than using the full first-order likelihood Eqs.~\ref{eq:loss}-\ref{eq:sigmatot}, and also did the analysis in the $\log(\go)-\gb$ rather than $\log(\gb)-\log(\go)$ plane (F. Lelli private communication).
The double power law fit has much higher maximum likelihood than the MOND IFs, outweighing its increased codelength due to its four free parameters. Although the RAR IF has complexity 9 it is not explicitly produced by ESR due to the constant ``1'' appearing,\footnote{This will be changed in a future version of ESR that performs ``integer snap'' of parameters where this reduces $L(D)$.} a generalised form in which this is replaced by a free parameter appears at rank 17, with a probability $4\times10^{10}$ times lower than the top-ranked function ($\Delta L(D)=24.5$). When $\theta_0\ne0$ the low-$\gb$ logarithmic slope $s_-$ of this function is 1 rather than 1/2, so it does not function as a MOND IF. We refer to it as the ``generalised RAR IF''.

The best ESR functions are clearly superior to the MOND functions or double power law. While the best metric for this is $L(D)$ (or equivalently $P(f)$),
other statistics lead to the same conclusion. There are many functions more accurate (lower $-\log(\mathcal{L})$) than even the double power law. Although the functions at rank 1-9 have more free parameters than the IFs this is more than compensated for by their greater accuracy: as an alternative metric, the Bayesian Information Criterion (BIC) of the rank 1 function is 108 lower than the Simple IF and even that of the rank 3 function with four free parameters is 94.5 lower, corresponding to a very strong preference. In agreement with~\citet{Paper_I, Paper_II, Paper_III}, when fitting the Simple IF + EFE we find that $e_\text{N}>0$ is clearly preferred, and recovers a value around the large-scale structure expectation. Although this function is significantly more accurate than any IF on its own, more than compensating for its additional free parameter ($\Delta\text{BIC}=-35$ compared to the Simple IF), it has a poor description length due to the large functional contribution. The complexity is 59 using our current basis set of operators, although this would fall to 45 if $\tanh$ were explicitly included. In general, the great improvement in accuracy and simplicity of the ESR functions demonstrates the advantage of this method over guessing functions ``by eye''.

\begin{table*}
  \begin{center}
    \begin{tabular}{l|c|c|c|c|c|c|c|c|c|c|c|}
      \hline
        \multirow{2}{*}{Rank} & \multirow{2}{*}{Function} & \multirow{2}{*}{Comp.} & \multirow{2}{*}{$P(f)$} & \multicolumn{4}{c|}{Parameters} & \multicolumn{4}{c|}{Description length}\\ 
\rule{0pt}{3ex}
      & & & & $\theta_0$ & $\theta_1$ & $\theta_2$ & $\theta_3$ & Resid.$^1$ & Func.$^2$ & Param.$^3$ & Total\\
      \hline
\rule{0pt}{3ex}
      1   &  $\theta_0 \left(| \theta_1 +x| ^{\theta_2}+x\right)$ & 9 & 9.3$\times 10^{-1}$ & 0.84 & -0.02 & 0.38 & --- & -1279.1 & 14.5 & 14.0 & -1250.6  \\
\rule{0pt}{3ex}
      2   & $\left| | \theta_1 | ^x+\theta_0 \right| ^{\theta_2} +x$ & 9 & 6.4$\times 10^{-2}$ & -0.99 & 0.64 & 0.36 & --- & -1279.9 & 12.5 & 19.6 & -1247.9 \\
\rule{0pt}{3ex}
    3   &  $| \theta_0 | ^{| \theta_1 -x| ^{\theta_2} -\theta_3}$ & 9 & 2.0$\times 10^{-3}$ & -1.4$\times 10^{2}$ & 0.02 & 0.14 & 0.89 & -1276.4 & 12.5 & 19.5 & -1244.4 \\
\rule{0pt}{3ex}
    4   &  $| \theta_0  (\theta_1 +x)| ^{\theta_2} +x$ & 9 & 1.4$\times 10^{-4}$ &  0.35 & -0.02 & 0.34 & --- & -1268.9 & 14.5 & 12.7 & -1241.7 \\
\rule{0pt}{3ex}
    5   &  $\left| \theta_0 -| \theta_1 -x| ^{\theta_2} \right| ^{\theta_3}$ & 9 & 1.0$\times 10^{-5}$  & -0.30 & 0.02 & 0.42 & 2.14 & -1271.1 & 12.5 & 19.5 & -1239.1 \\
\rule{0pt}{3ex}
    6   &  $\sqrt{x} \exp \left({\frac{| \theta_0 +x| ^{\theta_1}}{2}}\right) $ & 9 & 1.5$\times 10^{-9}$ & -0.02 & 0.36 & --- & --- & -1257.9 & 17.5 & 10.0 & -1230.3 \\
\rule{0pt}{3ex}
    7   &  $\left(\frac{| \theta_0 | ^x}{x}\right)^{\theta_1} +x$ & 9 & 2.4$\times 10^{-10}$ & 1.87 & -0.52 & --- & --- & -1250.6 & 14.5 & 7.6 & -1228.5 \\
\rule{0pt}{3ex}
    8   &  $\sqrt{| \theta_0 +x| }+\theta_1  x$ & 8 & 1.8$\times 10^{-10}$ & -1.8$\times 10^{-3}$ & 0.72 & --- & --- & -1245.6 & 12.9 & 4.5 & -1228.2  \\
\rule{0pt}{3ex}
    9   &  $\left| \theta_0 +\frac{1}{\sqrt[4]{x}}\right| ^{\theta_1} $ & 8 & 9.6$\times 10^{-11}$ & -0.22 & -2.14 & --- & --- & -1251.1 & 14.3 &  9.2 & -1227.6 \\
\rule{0pt}{3ex}
    10   &  $\left(\sqrt{x}+\frac{1}{x}\right)^{\theta_0}+x$ & 9 & 8.2$\times 10^{-11}$ & -0.53 &  --- &  --- & --- & -1248.3 & 16.1 & 4.8 & -1227.4 \\
\rule{0pt}{3ex}
      \vdots  & \vdots & \vdots & \vdots & \vdots & \vdots  & \vdots & \vdots & \vdots & \vdots & \vdots & \vdots\\
\rule{0pt}{3ex}
      17 & $x/(\exp(\theta_0) - |\theta_1|^{\sqrt{x}})$ & 9 & 2.2$\times 10^{-11}$ & 0.03 & 0.44 & --- & --- & -1250.9 & 17.5 & 7.3 & -1226.1  \\
      \hline
\rule{0pt}{3ex}
      --- & Double power law & 11 & 9.7$\times10^{-16}$ & 4.65 & 3.96 & 1.03 & 0.60 & -1252.3 & 17.7 & 18.5 & -1216.1\\
\rule{0pt}{3ex}
      --- & Simple IF & 10 & 5.5$\times10^{-25}$ & 1.11 & --- & --- & --- & -1217.3 & 18.6 & 3.9 & -1194.8\\
\rule{0pt}{3ex}
      --- & RAR IF & 9 & 6.7$\times10^{-26}$ & 1.13 & --- & --- & --- & -1212.8 & 16.1 & 3.9 & -1192.7\\
\rule{0pt}{3ex}
      --- & Simple IF + EFE & 59 & 5.0$\times10^{-69}$ & 1.16 & 6.8$\times10^{-3}$ & --- & --- & -1238.9 & 139.9 & 5.6 & -1093.4\\
\rule{0pt}{3ex}
      --- & Standard IF & 14 & 9$\times10^{-150}$ & 1.54 & --- & --- & --- & -939.5 & 27.9 & 4.1 & -907.5\\
      \hline
    \end{tabular}
    \begin{tabular}{c}
		$^1 - \log\mathcal{L} ( \hat{\bm{\theta}} ) $ \qquad\qquad $^2 k\log(n) + \sum_j \log(c_j)$ \qquad\qquad $^3 - \frac{p}{2} \log(3) + \sum_i^p (\log(I_{ii})^{1/2} + \log(|\hat{\theta}_i|))$ \
	\end{tabular}
  \caption{Top functions found by ESR applied to the SPARC data, ranked by total description length, compared to four MOND IFs and a double power law below the horizontal line. $x \equiv \gb / 10^{-10}$ m s$^{-2}$. We include also the ``generalised RAR IF'' at rank 17, although this does not have a deep-MOND limit. The IFs and double power law are not produced explicitly by our implementation of ESR and hence their ranks are unknown, although they are clearly worse than the best ESR functions in both description length and likelihood. For the Simple, RAR and Standard IFs the parameter is $g_0$, while for the ``Simple IF + EFE'' (Eq.~\ref{eq:efe}), the first is $g_0$ and the second $e_\text{N}$.}
  \label{tab:real_results}
  \end{center}
\end{table*}


In the top panel of Fig.~\ref{fig:ESR_results} we plot the best 20 functions on top of the SPARC data.
The functions are colour-coded by $P(f)$, with darker colouring indicating functions favoured by MDL. The top six functions have discontinuities in $s$ at $\gb \approx 0.02$. We have checked that this is not due to outlying points, does not invalidate the first-order approximation of Eq.~\ref{eq:loss}, and does not appear to map onto any local feature of the data. Instead it is likely due to the relative simplicity (i.e. complexity $\le9$) of the functions considered, as we discuss further in Sec.~\ref{sec:disc}. While such behaviour could be excluded by requiring no discontinuity in the derivative within the range of the data (or more generally weighting functions with an $s$-dependent prior), we see no principled reason to do so. The first function without a discontinuity is at rank 7, which has the Newtonian limit and $s_-=0.52$ at maximum likelihood. Although this is 9.3 $\sigma$ from $s_-=1/2$, and hence this equation cannot function as a pure-MOND IF, the EFE leads to the expectation that $s>1/2$ at low $\gb$ as discussed in more detail below. Several of the remaining highly ranked functions have similar quasi-MONDian behaviour.

While some of the best functions have MONDian limits around $\gb = 10^{\pm3}$ others do not, and in particular $s_- < 1/2$ is common. To explore this further we calculate $s_-$ and $s_+$ analytically for the top ten equations as a function of their free parameters, showing the results in Table~\ref{tab:limits_data}. For each of the functions where $s_-$ and/or $s_+$ depend on $\bm{\theta}$ (i.e. are not fixed by the functional form alone) we perform a Markov Chain Monte Carlo (MCMC) inference using the \texttt{numpyro} sampler \citep{phan2019composable, bingham2019pyro} with broad flat priors to constrain the parameters and hence derive the posterior predictive distributions of the limiting slopes. Fig.~\ref{fig:slopes} (left panel) shows the results for the top ten functions, using a dot to indicate a limit fixed by the functional form, a bar to show the 95 per cent confidence interval in cases where the slope depends on the parameters, and an arrowhead to indicate a limit outside of the plotting range (including $\pm\infty$). In many cases the 95\% confidence interval is extremely narrow.


Although it is $s_-$ and $s_+$ that directly relate to the MOND hypothesis, they require extrapolation far beyond the range of the data. To understand how the slopes behave near the limits of the SPARC data we also calculate $s$ at the minimum, $g_\text{bar, min}=8.32\times10^{-13} \: \text{ms}^{-2}$, and maximum, $g_\text{bar, max}=6.54\times10^{-9} \: \text{ms}^{-2}$, measured baryonic accelerations. These are plotted in cyan and magenta respectively in Fig.~\ref{fig:slopes}. At $g_\text{bar, max}$ the logarithmic slope is $\lesssim1$ for almost all functions, as expected for the Newtonian limit as only at $\gb\to\infty$ does $s$ become 1. However, we find that the $g_\text{bar,min}$ slopes of the top five functions are not $\sim1/2$ but actually $<0$ due to the aforementioned discontinuity. The remaining top functions have low-acceleration slopes typically slightly larger than 1/2. 

These results show that the SPARC data do not unambiguously favour $s_-=1/2$ and $s_+=1$. The requirement for an interpolating function to be MONDian is in fact even more stringent than this, since the coefficient of proportionality in the limiting high-$\gb$ power-law relation must be unity, i.e. $\go=\gb$. We find that among the functions in the top ten for which $s_+=1$, four have such a coefficient (at rank 2, 4, 7, 10) while for the rank 1 function this is $0.84\pm0.006$ and at rank 8 it is $0.72\pm0.01$, where the uncertainties are obtained by fitting the functions with MCMC. At low $\gb$ the coefficient in $\go\propto\gb^{1/2}$ should be $\sqrt{g_0}\approx1$. The only function with $s_-=1/2$ (rank 6) has the coefficient 1.12 (close to the 1.10 expected from the canonical $g_0=1.2$), while those further down with $s_\text{gbar,min}\approx1/2$ have a coefficient of 1. The relative simplicity of the functions we consider here should preference a coefficient of 1, and hence again it is not clear to what extent the data may be said to be MONDian. The double power law has the limits $\go=0.81\:\gb^{1.03}$ at high $\gb$ and $\go=1.57\:\gb^{0.60}$ at low $\gb$.

Next, we show in the left panel of Fig.~\ref{fig:Pareto} the separate Pareto fronts of description length and negative log-likelihood, with the second (``simplicity'') objective measured by functional complexity. Unlike the Pareto fronts produced by traditional SR algorithms, those of ESR are guaranteed to be optimal. $L(D)$ and $-\log(\mathcal{L})$ are minimised separately at each complexity, and have their minimum values over all complexity subtracted so that the globally best functions appear at 0. We show the MONDian functions and double power law as separate symbols, all of which we find to be strongly Pareto-dominated by the best ESR functions at lower complexity.
Note that while the ``knee'' of the Pareto front (where $L(D)$ or $-\log(\mathcal{L})$ turns over) would appear to be at complexity 6-7, there is a significant improvement in going from complexity 8 to 9. This cautions against automatically selecting functions at the knee (the default for example in \texttt{PySR}), and indicates that further improvement would likely be achievable by going beyond complexity 9. This is beyond the scope of the present work; we are content here to have discovered simple functional forms for the RAR surpassing any that have been considered heretofore. 

\subsection{Mock data}
\label{sec:mock_res}

The above results suggest a Newtonian limit ($\go\rightarrow\gb$ as $\gb\rightarrow\infty$) is somewhat favoured by the data while a deep-MOND limit ($\go\rightarrow\sqrt{g_0\:\gb}$ as $\gb\rightarrow0$) is questionable. However, given the limited dynamical range and significant uncertainties of the data it is unclear to what extent we should expect to find these limits even if the generating function were MONDian. In addition, the EFE would imply $s>1/2$ at low $\gb$. To investigate these issues we now apply ESR to the mock data of Sec.~\ref{sec:mock_method}.


\begin{table*}
  \begin{center}
  \setlength\tabcolsep{4.5pt}
    \begin{tabular}{l|c|c|c|c|c|c|c|c|c|c|c|}
      \hline
        \multirow{2}{*}{Rank} & \multirow{2}{*}{Function} & \multirow{2}{*}{Comp.} & \multirow{2}{*}{$P(f)$} & \multicolumn{4}{c|}{Parameters} & \multicolumn{4}{c|}{Description length}\\ 
\rule{0pt}{3ex}
      & & & & $\theta_0$ & $\theta_1$ & $\theta_2$ & $\theta_3$ & Resid.$^1$ & Func.$^2$ & Param.$^3$ & Total\\
      \hline
\rule{0pt}{3ex}
      1   & $\theta_0+\theta_1 x+\sqrt{x}$ & 8 & 5.6$\times 10^{-1}$ & 9.1$\times 10^{-3}$ & 0.63 & --- & --- & -2045.2 & 12.9 & 4.9 & -2027.4  \\
\rule{0pt}{3ex}
      2   & $\sqrt{| \theta_0+x| }+\theta_1 x$ & 8 & 2.8$\times 10^{-1}$ & 3.0$\times 10^{-3}$ & 0.64 & --- & --- & -2044.4 & 12.9 & 4.8 & -2026.7 \\
\rule{0pt}{3ex}
    3   & $\theta_0 x+x^{\theta_1}$ & 7 & 8.2$\times 10^{-2}$ & 0.64 & 0.49 & --- & --- & -2045.2 & 11.3 & 8.5 & -2025.5 \\
\rule{0pt}{3ex}
    4   & $\sqrt{x} \exp\left({\frac{x^{\theta_0}}{2}}\right)$ & 7 & 3.5$\times 10^{-2}$ & 0.36 & --- & --- & --- & -2040.7 & 12.5 & 3.5 & -2024.7 \\
\rule{0pt}{3ex}
    5   & $(\theta_0+x) \left(\theta_1+\frac{1}{\sqrt{x}}\right)$ & 9 & 1.1$\times 10^{-2}$ & 1.3$\times 10^{-3}$ & 0.64 &  --- & --- & -2044.5 & 16.1 & 4.8 & -2023.5  \\
\rule{0pt}{3ex}
    6   & $\frac{1}{\sqrt{\left| \theta_0+\frac{1}{x}\right| }}+x$ & 8 & 8.8$\times 10^{-3}$& 1.74 & --- & --- & --- & -2038.5 &  12.9 &  2.3 & -2023.3  \\
\rule{0pt}{3ex}
    7   & $(x | \theta_0| )^{(x | \theta_1| )^{\theta_2}}$ & 9 & 3.1$\times 10^{-3}$ & -2.09 & -1.4$\times 10^{-4}$ & 0.04  & --- & -2045.3 & 12.5 & 10.6 & -2022.2 \\
\rule{0pt}{3ex}
    8   & $\theta_0 x+| \theta_1+x| ^{\theta_2}$ & 9 & 2.4$\times 10^{-3}$ & 0.64 & 1.4$\times 10^{-3}$ & 0.49 & --- & -2045.4 &  14.5  & 8.9 & -2022.0 \\
\rule{0pt}{3ex}
    9   & $x \left(| \theta_0-x| ^{\theta_1}-\theta_2\right)$ & 9 & 2.3$\times 10^{-3}$ & 1.2$\times 10^{-3}$ & -0.51 & -0.64 & --- & -2045.3 & 14.5 &  8.9 & -2021.9 \\
\rule{0pt}{3ex}
    10   & $(\theta_0-x) \left(\theta_1-x^{\theta_2}\right)$ & 9 & 2.2$\times 10^{-3}$ & -6.5$\times 10^{-4}$ & -0.64  & -0.51 & --- & -2045.4 &  14.5  &  9.0  & -2021.9 \\
\rule{0pt}{3ex}
      \vdots  & \vdots & \vdots & \vdots & \vdots & \vdots  & \vdots & \vdots & \vdots & \vdots & \vdots & \vdots\\
\rule{0pt}{3ex}
      27 & $x/(\exp(\theta_0) - \exp(-\sqrt{x}))$ & 9 & 3.2$\times10^{-4}$ & -0.01 & --- & --- & --- & -2039.3 & 17.5 & 1.9 & -2020.0\\
\rule{0pt}{3ex}
      \vdots  & \vdots & \vdots & \vdots & \vdots & \vdots  & \vdots & \vdots & \vdots & \vdots & \vdots & \vdots\\
\rule{0pt}{3ex}
      41 & $x/(\exp(\theta_0) - |\theta_1|^{\sqrt{x}})$ & 9 & 1.1$\times10^{-4}$ & -5.0$\times10^{-3}$ & 0.38 & --- & --- & -2042.1 & 17.5 & 5.7 & -2018.9\\
    \hline
\rule{0pt}{3ex}
      --- & RAR IF & 9 & 1.0$\times10^{-3}$ & 1.14 & --- & --- & --- & -2041.1 & 16.1 & 3.9 & -2021.1\\
\rule{0pt}{3ex}
      --- & Double power law & 11 & 3.4$\times10^{-8}$ & 1.25 & 1.47 & 0.90 & 0.54 & -2047.2 & 17.7 & 18.7 & -2010.8\\
\rule{0pt}{3ex}
      --- & Simple IF & 10 & 2.8$\times10^{-11}$ & 1.12 & --- & --- & --- & -2026.2 & 18.6 & 3.9 & -2003.7\\
\rule{0pt}{3ex}
      --- & Standard IF & 14 & 2.9$\times10^{-55}$ & 1.54 & --- & --- & --- & -1934.4 & 27.9 & 4.1 & -1902.4\\
\rule{0pt}{3ex}
      --- & Simple IF + EFE & 59 & 5.9$\times10^{-64}$ & 1.12 & 0 & --- & --- & -2026.2 & 139.9 & 3.9 & -1882.4\\
      \hline
    \end{tabular}
    \begin{tabular}{c}
		$^1 - \log\mathcal{L} ( \hat{\bm{\theta}} ) $ \qquad\qquad $^2 k\log(n) + \sum_j \log(c_j)$ \qquad\qquad $^3 - \frac{p}{2} \log(3) + \sum_i^p (\log(I_{ii})^{1/2} + \log(|\hat{\theta}_i|))$ \
	\end{tabular}
  \caption{As Table~\ref{tab:real_results} but for the RAR IF mock data. We find the generalised RAR IF at rank 41, and a closely related function at rank 27 in which the free parameter appears in the other term in the denominator. For this dataset the RAR IF itself is superior to both of these modified forms, as would be expected given that it generated the data.}
  \label{tab:mock_results}
  \end{center}
\end{table*}

\subsubsection{RAR IF generating function}

Table~\ref{tab:mock_results} shows the best functions found by ESR for the RAR IF mock data, along with the results for the IFs and double power law. The RAR IF is by construction a good fit to this data, but there are 15 functions with lower $L(D)$, including several at lower complexity. This indicates that the characteristics of the SPARC data (dynamic range and uncertainties) are insufficient to pick out the true generating function: ESR prefers simpler functions which may achieve slightly higher likelihoods. Since the $-\log(\mathcal{L})$ term in $L(D)$ becomes dominant at large dataset size, simply increasing the number of (mock) observations with otherwise identical properties would not be sufficient to push the generating function to the top of the list.
For this dataset we find that the generalised RAR IF, appearing at rank 41, has higher $L(D)$ than the RAR IF despite slightly higher likelihood, a success of MDL's penalisation of more complex functions. Note that the best-fit generalised RAR IF is $x/(0.995-\exp(-\sqrt{x/1.068}))$, somewhat offset from the ground-truth values \{1, 1.127\}. This results from a combination of the limited dataset size (introducing random noise) and a small bias in the maximum-likelihood estimator which we discuss further in Appendix~\ref{sec:app_loss}. At rank 27 we find a close cousin of the RAR IF in which the ``1'' is free and $g_0$ is pinned to 1. This performs only slightly worse than the RAR IF itself because the true $g_0$ is close to 1.

The highest ranked ESR function is better by $\Delta L(D)=6.3$ than the RAR IF. Although the relative probability between the best function and the best RAR-like function is smaller than in the observations ($\sim1600$ compared to $4\times10^{10}$), it is interesting that the best RAR-like function appears further up the list in the real data (rank 17 vs 27). The double power law is disfavoured relative to the RAR IF and many ESR functions despite having the highest likelihood shown in the table, another success of the complexity penalisation. There are a few functions overall with lower $-\log(\mathcal{L})$, the lowest being $-2048.0$ at rank 51 ($(\theta_0 + |\theta_1 + \theta_2/x|^{1/2})^{-1}$, $L(D)=-2018.4$). Also as expected, the Simple IF + EFE has $e_\text{N}$ snapped to 0 and hence behaves identically to the Simple IF, albeit with larger functional complexity.

The top 20 ESR functions for the RAR IF mock are overplotted on that data in the middle panel of Fig.~\ref{fig:ESR_results}. We find a slightly reduced spread in $s$ at both the high-$\gb$ and low-$\gb$ ends compared to the real data, without any discontinuities within the data range. However there is still significant uncertainty beyond the range of the data. This is quantified in the middle panel of Fig.~\ref{fig:slopes}, where the top ten functions are all observed to have slopes of approximately $1/2$ at $g_\text{bar,min}$ and $1$ at $g_\text{bar,max}$, although only in two cases is $s_-=1/2$: for the others it is lower. This indicates that constraining the slope to be near $1/2$ at $g_\text{bar,min}$ is insufficient to conclude that $s_-$ takes a similar value, at least up to complexity 9. One would need to reduce the uncertainties, or, preferably, lower $g_\text{bar,min}$. That this applies to a lesser extent at high-$\gb$ is shown by the functions at rank 4 and 7 with $s_+=\infty$.

Adding to the conclusion that the mock data characteristics are insufficient to pick out a MONDian generating function, we find that the coefficient in $\go \propto \gb$ at high $\gb$ is only unity for one of the top-10 functions for which $s_+=1$ (at rank 6). For all the others it is 0.64, with the exception of that at rank 1 where it is 0.63. These values have uncertainties $\sim0.003$ when constrained by MCMC, and vary by $\sim0.02$ over mock datasets differing only in the random seed. Thus the best functions almost always fail to recover the Newtonian limit even when it is the truth, presumably due to an insufficient $g_\text{bar,max}$. The origin of 0.64 is unclear, but presumably results from the way the lower-$\gb$ behaviour is filtered through the forms of the functions found to be optimal. In cases where $s_-=1/2$, the coefficient of proportionality is 1 (to be compared to $\sqrt{g_0}$ in MOND), and the double power law limits are $\go=1.20\:\gb^{0.90}$ at high $\gb$ and $\go=1.30\:\gb^{0.54}$ at low $\gb$.

The middle panel of Fig.~\ref{fig:Pareto} shows the Pareto front for these data. We find more smooth behaviour than for the real data, with the optimum solution achieved already by complexity 8, reflecting the relatively simple nature of the generating function. This shows that if the RAR IF was generating the real data (and our likelihood and mock data generation method were accurate), we would have achieved the $L(D)$ minimum on those data too. However, the MONDian functions (including the RAR IF itself) and double power law are Pareto-dominated by the ESR results even on these mock data, showing that one should not expect to be able to recover unambiguously even this simplest of MOND generating functions.

\begin{table*}
  \begin{center}
    \begin{tabular}{l|c|c|c|c|c|c|c|c|c|c|c|}
      \hline
        \multirow{2}{*}{Rank} & \multirow{2}{*}{Function} & \multirow{2}{*}{Comp.} & \multirow{2}{*}{$P(f)$} & \multicolumn{4}{c|}{Parameters} & \multicolumn{4}{c|}{Description length}\\ 
\rule{0pt}{3ex}
      & & & & $\theta_0$ & $\theta_1$ & $\theta_2$ & $\theta_3$ & Resid.$^1$ & Func.$^2$ & Param.$^3$ & Total\\
      \hline
\rule{0pt}{3ex}
      1   & $\theta_0+\sqrt{x^2+2 x}$ & 9 & 8.9$\times 10^{-1}$ & -0.06 & --- & --- & --- & -2017.7 & 14.5 & 3.1 & -2000.0 \\
\rule{0pt}{3ex}
      2   & $\theta_0+\sqrt{x | \theta_1 +x| }$ & 8 & 9.3$\times 10^{-2}$ & -0.06 & 1.97 & --- & --- & -2017.9 & 12.9 & 7.3 & -1997.8 \\
\rule{0pt}{3ex}
    3   & $-| \theta_0| ^{\sqrt{x}}+\theta_1 +x$ & 8 & 5.6$\times 10^{-3}$ & 0.26 & 0.95 & --- & --- & -2017.9 & 12.9 &  10.1 & -1995.0 \\
\rule{0pt}{3ex}
    4   & $(\theta_0-x) \left(\theta_1 -x^{\theta_2}\right)$ & 9 & 3.3$\times 10^{-3}$ & 3.1$\times 10^{-3}$  & -0.71 & -0.53 & --- & -2019.7 & 14.5 & 10.7 & -1994.4 \\
\rule{0pt}{3ex}
    5   & $x^{\theta_0}-\theta_1  (\theta_2-x)$ & 9 & 2.4$\times 10^{-3}$ & 0.39 & 0.79 & 0.12 & --- & -2020.9 & 14.5 & 12.3 & -1994.1 \\
\rule{0pt}{3ex}
    6   & $| \theta_0-x| ^{\theta_1} -\theta_2 x$ & 9 & 2.0$\times 10^{-3}$ & 5.5$\times 10^{-3}$ & 0.48 & -0.71 & --- & -2019.1 &  14.5 & 10.6 & -1994.0\\
\rule{0pt}{3ex}
    7   & $x | \theta_0| ^{-| \theta_1 | ^{x^{\theta_2}}}$ & 9 & 1.7$\times 10^{-3}$ & 0.04 & -0.16 & 0.33 & --- & -2018.1 & 12.5 & 11.9  & -1993.8\\
\rule{0pt}{3ex}
    8   & $x \left(\theta_0 + |\theta_1 +x| ^{\theta_2}\right)$ & 9 & 1.5$\times 10^{-3}$ & 0.71 & 0.01 & -0.53 & --- & -2018.7 & 14.5 & 10.6 & -1993.7 \\
\rule{0pt}{3ex}
    9   & $| \theta_0| ^{| \theta_1 | ^{x^{\theta_2}}}+x$ & 9 & 6.5$\times 10^{-4}$ & 7.0$\times 10^{-6}$ & 0.03 & 0.17 & --- & -2016.7 & 12.5 & 11.4 & -1992.8 \\
\rule{0pt}{3ex}
    10   & $\exp\left({\theta_0-\frac{1}{\sqrt[4]{x}}}\right)+x$ & 9 & 5.5$\times 10^{-4}$ & 0.57 & --- & --- & --- & -2014.0 & 17.5 & 3.9 & -1992.6 \\
\rule{0pt}{3ex}
      \vdots  & \vdots & \vdots & \vdots & \vdots & \vdots  & \vdots & \vdots & \vdots & \vdots & \vdots & \vdots\\
\rule{0pt}{3ex}
      21 & $x/(\exp(\theta_0) - |\theta_1|^{\sqrt{x}})$ & 9 & 1.8$\times10^{-5}$ & 0.03 & 0.44 & --- & --- & -2014.2 & 17.5 & 7.4 & -1989.3 \\
      \hline
\rule{0pt}{3ex}
      --- & Double power law & 11 & 3.4$\times10^{-11}$ & 3.53 & 3.31 & 0.98 & 0.60 & -2012.3 & 17.7 & 18.6 & -1976.0\\
\rule{0pt}{3ex}
      --- & Simple IF & 10 & 1.2$\times10^{-22}$ & 1.11 & --- & --- & --- & -1972.1 & 18.6 & 3.9 & -1949.6\\
\rule{0pt}{3ex}
      --- & RAR IF & 9 & 7.0$\times10^{-24}$ & 1.13 & --- & --- & --- & -1966.9 & 16.1 & 3.9 & -1946.8\\
\rule{0pt}{3ex}
      --- & Simple IF + EFE & 59 & 3.8$\times10^{-57}$ & 1.19 & 8.6$\times10^{-3}$ & --- & --- & -2016.0 & 139.9 & 5.9 & -1870.2\\
\rule{0pt}{3ex}
      --- & Standard IF & 14 & 2$\times10^{-141}$ & 1.54 & --- & --- & --- & -1708.3 & 27.9 & 4.1 & -1676.3\\
      \hline
    \end{tabular}
    \begin{tabular}{c}
		$^1 - \log\mathcal{L} ( \hat{\bm{\theta}} ) $ \qquad\qquad $^2 k\log(n) + \sum_j \log(c_j)$ \qquad\qquad $^3 - \frac{p}{2} \log(3) + \sum_i^p (\log(I_{ii})^{1/2} + \log(|\hat{\theta}_i|))$ \
	\end{tabular}
  \caption{As Table~\ref{tab:real_results} but for the Simple IF + EFE mock data.}
  \label{tab:mock_results_efe}
  \end{center}
\end{table*}

\subsubsection{Simple IF + EFE generating function}

Analogous results for the mock data generated using the Simple IF with inclusion of the EFE are shown in Table~\ref{tab:mock_results_efe} and the bottom/right panels of Figs.~\ref{fig:ESR_results}-\ref{fig:Pareto}. This dataset behaves more similarly to the real data in terms of the relative ordering of the IFs, double power law and ESR functions, including the generalised RAR IF. Indeed, the best-fit parameters of all three non-EFE IFs are identical to the SPARC data to two decimal places, while those of the generalised RAR IF and the high-$\gb$ slope of the double power law are the same to one. Here the IFs provide a significantly worse compression of the data than the best ESR functions, and the double power law also performs relatively poorly due to the curvature at low $\gb$ (see Fig.~\ref{fig:IFs}). There is again a small bias between the maximum-likelihood (1.19 and 8.56$\times10^{-3}$) and true (1.2 and 1$\times10^{-2}$) $g_0$ and $e_\text{N}$ values for the Simple IF + EFE fit. Although this function has among the highest likelihoods achievable by ESR up to complexity 9, its functional complexity makes it a poor compression of its own SPARC-like mock data. This reinforces the conclusion that the characteristics of these data are insufficient to identify a MONDian generating function: this one in particular would require far more data than the RAR IF to be favoured by MDL.

For the Simple IF + EFE mock all top-10 functions have $s_+=1$ and $s(g_\text{bar,max})>0.9$. However, only six of them recover the true Newtonian limit $\go=\gb$: the others find $\go=0.71\:\gb$ or $\go=0.79\:\gb$, again with sub-percent uncertainty from MCMC. Thus under this model too one would not expect the Newtonian limit to be identified robustly. While $s(g_\text{bar,min})>1/2$, indicating the significant impact of the EFE, $s_-$ is typically 0 as opposed to 1 as expected from Eq.~\ref{eq:efe}. Thus $g_\text{bar,min}$ is too high to constrain $s_-$ reliably, although this may also be a reflection of the relative simplicity of the functions we consider. The double power law limits are $\go=0.96\:\gb^{0.98}$ at high $\gb$ and $\go=1.55\:\gb^{0.60}$ at low $\gb$. The Pareto front indicates this dataset to be somewhat more complex than the RAR IF mock, as $L(D)$ continues to fall to complexity 9, although the smoothness shows it to be simpler than the real data.

\vspace{3mm}

\noindent All functions considered achieve considerably higher likelihood on the mock datasets than the real data, showing that the mocks are simpler. This could be because the data do not conform to the MOND expectation---one would expect any given $\go=f(\gb)$ to be relatively inaccurate in a chaotic $\Lambda$CDM galaxy formation scenario---or because the model for scattering the mock data points is overly simplistic. This is discussed further in Sec.~\ref{sec:disc}. Relatedly, the $P(f)$ values of the top functions are very closely spaced in the mock datasets, indicating that there is little to distinguish them. On the contrary, on the real data the integrated probability of all functions besides the top five is $\lesssim$10$^{-8}$, suggesting that these functions perhaps ought not to be considered at all. The limiting slopes of the top ten equations as a function of the parameters can be found in Tables~\ref{tab:limits_rar} and~\ref{tab:limits_efe} for the RAR and EFE mocks respectively.

\section{Discussion}
\label{sec:disc}



Our main conclusion is that the SPARC data are insufficient to determine robustly the limiting behaviour of the RAR, and hence cannot verify or refute the MOND hypothesis. This is reached by studying mock data generated by MOND; in particular, generating data according to the RAR IF, not only are we unable to identify that as the generating function, but, more seriously, we cannot reconstruct $s_-=1/2$. At the high-$\gb$ end, the logarithmic slope of the Newtonian limit ($s_+=1$) is typically well recovered, although the coefficient of proportionality in $\go\propto\gb$ is not: in the RAR IF mock data this takes a values $\sim0.64$ far more often than $1$.

Improving this situation requires increasing the dynamical range of the RAR. At the low-$\gb$ end this may be achieved by studying ultra-diffuse galaxies (e.g.~\citealt{Freundlich}), or local dwarf spheroidals (e.g.~\citealt{McGaugh_Wolf, McGaugh_Milgrom}), some of which seem seem to indicate $s_-\approx0$, as found by many of the best ESR functions \citep{RAR}. Alternatively, one may attempt to probe the outer regions of galaxies including the Milky Way (e.g.~\citealt{Oman}). Particularly promising for a large gain is to use stacked weak lensing to probe galaxy outskirts that would have insufficient signal-to-noise on an individual-object basis \citep{Brouwer}. This appears to indicate $s_-\approx1/2$. Increasing $g_\text{bar,max}$ requires probing the central regions of high-mass ellipticals, as well as groups and clusters of galaxies \citep{ellipticals_1, ellipticals_2, groups, clusters_1, clusters_2, clusters_3}. Such data already exists and may readily be folded into our framework to increase its constraining power. A smaller information gain may be achieved by reducing the uncertainties in $\gb$ and $\go$: in the limit of no uncertainty any generating function will be assigned $P(f)=1$ by MDL. By generating mock data with different characteristics one could ascertain the requirements for various features of the functional form to be unambiguously determined.

It is likely that there exist functions at higher complexity superior to those of Tables~\ref{tab:real_results}-\ref{tab:mock_results_efe}, especially for the real data where $L(D)$ drops significantly from complexity 8 to 9. While uncovering lower-$L(D)$ functions at complexity $>$9 may update the optimal limiting behaviours of the functional form of the RAR, and hence its compatibility with MOND, it cannot compromise our discovery of simpler and more accurate functions than the IFs and double power law.
Indeed, the fact that the (or at least \emph{a}) knee of the Pareto front is reached around complexity 7 in the left panel of Fig.~\ref{fig:Pareto} shows that such functions already offer a powerful compression, and the commonalities between the top functions at complexity $\lesssim$9 suggests that similar features are likely to be present in more complex functions also.

Discovering such functions would likely be computationally prohibitive for the ESR algorithm, and thus a stochastic search (e.g. using a genetic algorithm) may be required. This search may be seeded by the ESR functions: the fact that many of the best-fitting functions have similar features (such as $\go\approx0.64\:\gb$ as $\gb\rightarrow\infty$ for the RAR IF mock) suggests these may be useful for higher-complexity functions also.
Thus ESR may be used to validate the underlying assumption of stochastic searches---that there exist features of functions responsible for their fitness---and the identification of these features may be useful for tuning hyperparameters. It would also be possible to combine ESR with deterministic symbolic regression algorithms (e.g. \citealt{Worm, Dome, Kammerer_2021}) to search systematically the neighbourhood of good functions towards higher complexity.

Our best functions on the real data have a discontinuity in $s$ around $\gb=0.02$. This is likely due to the limited complexity of the equations we consider: a cusp is the simplest way of changing $s$ sharply. It is probable that the optimal functions at higher complexity will have a smoothed form of this behaviour in which $s$ does not become negative and may not tend to 0. We therefore doubt that the $s_-$ and $s(g_\text{bar,min})$ values of the best functions in Table~\ref{tab:real_results} are robust. One could attempt to construct more complex functions inspired by the ESR results with similar but not discontinuous behaviour and calculate their $-\log(\mathcal{L})$ and $L(D)$ separately, or feed them into a genetic algorithm as mentioned above. On the other hand, below complexity 9, there is only a single low description length function that is discontinuous, the third best function at complexity 8 ($P(f)=4.2\times10^{-11}$). The best functions at lower complexity more frequently have $s_-=1/2$ and $s_+=1$, although again they rarely satisfy $\go=\gb$ as $\gb\rightarrow\infty$. For example, the top function at complexity 6---marking the (first) knee of the $L(D)$ Pareto front in Fig.~\ref{fig:Pareto}---is $\go=0.70\:\gb + \sqrt{\gb}$, exhibiting similar high-$\gb$ behaviour to the 1$^\text{st}$- and 8$^\text{th}$-ranked functions overall.

To generate the mock data we assumed that all $\gb$ values are uncorrelated. While this is likely true between galaxies, it is not within a single galaxy because the uncertainty in $\gb$ is dominated by the mass-to-light ratio, a global galaxy parameter in the simplest approximation. This may be seen from the data in Fig.~\ref{fig:IFs}, where lines of points (e.g. scattering low around $\gb=2$ or high around $\gb=10$) are all from the same galaxy. A more robust procedure may be to generate $\Upsilon$ values for each mock galaxy by randomly drawing from their priors, use this to transform $\gb$ and then add any other, random sources of noise (e.g. from the uncertainty in 3.6 $\mu$m luminosity). By enhancing inter-galaxy variations this may increase the complexity of the mock datasets, moving their ESR results towards those of the SPARC data. Alternatively, one may fit each galaxy separately to assess compatibility of their individual RARs (analogously to \citealt{Li} but not just for the RAR IF). The assumption of uncorrelated data points is also present in our likelihood, as discussed further in Appendix~\ref{sec:app_loss}. A complete analysis would infer $D$, $i$, $L_{3.6}$ and $\Upsilon$ for each SPARC galaxy along with the parameters of the function being fitted.

We have assumed no intrinsic scatter in the RAR, such that all deviations from the hypothetical functional expectation must come from the observational uncertainties. While this is expected in MOND, in $\Lambda$CDM the complex process of galaxy formation would lead to a significant and parameter-dependent effective intrinsic scatter \citep{Desmond_MDAR}. Even the EFE would introduce some scatter due to galaxy-by-galaxy variation in $g_\text{ex}$~\citep{Paper_II}. It would be straightforward to add this (in some direction on the RAR plane) as an additional free parameter of all functions, which would alter the results. MDL naturally penalises the addition of this parameter, allowing one to determine whether it is justified for any given function. This would provide further evidence concerning the optimality of MOND by assessing the extent to which the data implies law-like modified gravity behaviour.

Our current implementation of MDL treats the parameter values as part of the model and chooses
them to maximise the likelihood. An alternative would be to treat the hypothesis in Eq.~\ref{eq:length_final} as the functional form alone, assigning codelengths and probabilities to functions regardless of their parameter values. In a Bayesian formulation this corresponds to marginalising over the parameters, and enables a simpler one-part coding scheme where the description length is simply the negative logarithm of the model evidence including any functional prior. An even higher-level approach would be to group functions into sets with specific properties, e.g. limiting behaviour. This would enable calculation of the posterior predictive distribution of any feature of the functional representation of the dataset, and hence enable model comparison at any level of generality.

The relative simplicity of the RAR and conformity to the Newtonian and deep-MOND limits are the key differences between the expectations of MOND and the more chaotic galaxy formation scenario of $\Lambda$CDM: it is only under the simpler scenario that one would \emph{expect} to find a simple $\go = \mathcal{F}(\gb)$. While our results are therefore not particularly supportive of the MOND hypothesis, this is not to say either that the data could not plausibly have been generated by MOND or that it could plausibly have been generated under another hypothesis, as only MOND currently has sufficient predictivity for a test of this precision. We look to future SR studies with more data to establish the functional form of the RAR---if it exists---definitively.






\section{Conclusion}
\label{sec:conc}

The radial acceleration relation (RAR) has become central to debates about the mass discrepancy problem on astrophysical scales. Its tightness and regularity have been used to argue for a violation of Newtonian gravity in accordance with Modified Newtonian Dynamics (MOND), but the functions used to fit the data have been constructed to conform to this theory. As the first detailed application of the brand-new technique of Exhaustive Symbolic Regression, we rank objectively \emph{all} simple functions in terms of their aptitude for describing the SPARC RAR. We employ the minimum description length principle to trade accuracy with simplicity and hence perform model selection, and calibrate our method on mock MOND data generated both with and without the external field effect (EFE). Our conclusions are as follows:
\begin{itemize}
    \item ESR discovers functions which are better descriptions, in both accuracy and simplicity and for both observed and simulated data, than MOND functions or a double power law.
    \item While the majority of best-fitting functions on the SPARC data recover $\go\propto\gb$ at high accelerations, not all have a best-fit coefficient of proportionality near unity. Thus the Newtonian limit is not clearly evidenced.
    \item The SPARC data do not prefer functions with the deep-MOND limit of $\go \propto \sqrt{\gb}$ as $\gb \to 0$. Instead, we find that functions with $\gb\to {\rm const}$ typically compress the data more efficiently, albeit with considerable uncertainty.
    \item SPARC-like mock data generated assuming the MONDian RAR interpolating function do not unambiguously recover that function. Moreover, many of the best functions for those mock data have $\go\approx0.64\:\gb$ rather than $\go=\gb$ at high $\gb$, and most do not have a deep-MOND limit at all.
    \item The EFE in AQUAL greatly increases the logarithmic slope of the best-fitting functions at the low-$\gb$ end of the data, but does not appreciably impact the limiting slope at $\gb \to 0$. Incorporating the EFE in the mock data produces more generally similar results to the real data, so our analysis (within the MOND paradigm) hints at it.
    \item We conclude that the data have too small a dynamic range (and too large uncertainties) to unambiguously favour MOND even if it is in fact generating the data. The SPARC RAR alone, therefore, does not supporting that theory unambiguously. The best prospect for improving this situation is to increase the acceleration range of the data, e.g. using stacked weak lensing at low $\gb$ and groups and clusters at high $\gb$.
    \item Our results are a function of the maximum complexity of equation considered. Future symbolic regression algorithms---exhaustive or non-exhaustive---will reach the true description length minimum and hence uncover the optimal functional representation of the RAR and determine whether the relation implies novel law-like gravitational behaviour.
\end{itemize}

Exhaustive Symbolic Regression provides for the first time a guaranteed complete search through functional parameter space, making it the ideal tool to determine the analytic form of observed relations, extract physics from data theory-agnostically, and create fitting functions. We make the ESR and RAR codes, full function sets and the best 50 functions for each dataset we consider publicly available to facilitate future applications.

\section{Data availability}

The code and data associated with ESR and its application to the RAR are released at \esrgithub \ and in \citet{ESR_data}.
The SPARC data is available at \url{http://astroweb.cwru.edu/SPARC}.
Other data may be shared on request to the corresponding authors.

\section*{Acknowledgements}


We thank Kyu-Hyun Chae, Andrei Constantin, Miles Cranmer, Mario Figueiredo, Gianluca Gregori, Thomas Harvey, Mark Kotanchek, Federico Lelli, Stacy McGaugh, Andre Lukas, Richard Stiskalek and Tariq Yasin for useful inputs and discussion. 

HD is supported by a Royal Society University Research Fellowship (grant no. 211046).
DJB is supported by the Simons Collaboration on ``Learning the Universe'' and was supported by STFC and Oriel College, Oxford.
PGF acknowledges support from European Research Council Grant No: 693024 and the Beecroft Trust.

This project has received funding from the European Research Council (ERC) under the European Union's Horizon 2020 research and innovation programme (grant agreement No 693024).

This work used the DiRAC Complexity and DiRAC@Durham facilities, operated by the University of Leicester IT Services and Institute for Computational Cosmology, which form part of the STFC DiRAC HPC Facility (www.dirac.ac.uk). This equipment is funded by BIS National E-Infrastructure capital grants ST/K000373/1, ST/P002293/1, ST/R002371/1 and ST/S002502/1, STFC DiRAC Operations grant ST/K0003259/1, and Durham University and STFC operations grant ST/R000832/1. DiRAC is part of the National E-Infrastructure.

For the purpose of open access, the authors have applied a Creative Commons Attribution (CC BY) licence to any Author Accepted Manuscript version arising.

\bibliographystyle{mnras}
\bibliography{references}

\appendix

\section{A note on likelihoods}
\label{sec:app_loss}

\subsection{Correlation of measurements and their uncertainties}
\label{sec:app_corr}

The likelihood of Sec.~\ref{sec:loss} treats the uncertainties induced by $\Upsilon$, $D$ and $i$ as statistical and uncorrelated between points. This is clearly incorrect: a scattering up of $\Upsilon$, for example, causes a coherent increase in $\gb$ across the rotation curve, generating off-diagonal elements in the covariance matrix. A better approach is therefore to calculate the covariance matrix via Monte Carlo. For a given galaxy, one would independently sample $\Ug, \Ud, \Ub, D$ and $i$ many times from their assumed-Gaussian prior distributions. One would then generate the corresponding $\mathbf{\go}$ and $\mathbf{\gb}$ values (now vectors across the rotation curve of a given galaxy), scatter them by their statistical uncertainty ($\mathbf{\gb}$ has only the small $\delta L_{3.6}$ term), and calculate the covariance matrices $\mathbf{\mathsf{\Sigma}}_\text{obs}$ and $\mathbf{\mathsf{\Sigma}}_\text{bar}$.

Let us define $u \equiv \log\left(\gb\right)$ and $v \equiv \log \left( \go \right)$, such that $\bm{u}$ and $\bm{v}$ are vectors containing $u$ and $v$ for all galaxies at all measured points along the rotation curve. We denote the ``true'' values with a superscript $\text{t}$ and the observed values without a superscript. We assume that $\bm{u}^\text{t}$ and $\bm{v}^\text{t}$ are drawn from multivariate Gaussian distributions with covariance matrices $\mathsf{\Sigma}_u$ and $\mathsf{\Sigma}_v$, respectively. Since $\bm{u}^\text{t}$ and $\bm{v}^\text{t}$ are statistically independent random variables for fixed $\bm{u}$ and $\bm{v}$, the likelihood of a given set of points in the $\left(\bm{u}^\text{t}, \bm{v}^\text{t} \right)$ plane would then be
\begin{equation}
    \label{eq:uv_likelihood}
    \begin{split}
        \mathcal{L} \left( \bm{u}, \bm{v} \right) &= 
        \frac{1}{\sqrt{ |2 \pi \mathbf{\mathsf{\Sigma}}_u|}}\frac{1}{\sqrt{| 2 \pi \mathbf{\mathsf{\Sigma}}_v|}}  \\
        &\times  \exp \left( -\frac{1}{2} \left(\bm{u}^\text{t}-\bm{u}\right)^{\rm T} \mathbf{\mathsf{\Sigma}}^{-1}_u \left(\bm{u}^\text{t}-\bm{u} \right) \right) \\
        &\times \exp \left( -\frac{1}{2} \left(\bm{v}^\text{t}-\bm{v} \right)^{\rm T}  \mathbf{\mathsf{\Sigma}}^{-1}_v \left(\bm{v}^\text{t}-\bm{v} \right) \right).
    \end{split}
\end{equation}
The equivalent of Eq.~\ref{eq:loss} for a vector-valued function $\bm{f}$, $\bm{v} = \bm{f}\left( \bm{u} \right)$, is then obtained by marginalising over $\bm{u}^\text{t}$,
\begin{equation}
    \label{eq:v_likelihood}
    \mathcal{L} \left( \bm{v} \right) = \int \mathcal{L} \left( \bm{u}^\text{t}, \bm{f}\left( \bm{u}^\text{t} \right) \right) \dd \bm{u}^\text{t}.
\end{equation}
Taylor expanding $\bm{f}\left(\bm{u}^\text{t}\right)$ about $\bm{u}$, 
\begin{equation}
    \label{eq:taylor}
   \bm{f}\left(\bm{u}^\text{t} \right) \approx 
   \bm{f} \left(\bm{u} \right) +
    \mathbf{\mathsf{D}} \left( \bm{u}^\text{t}-\bm{u} \right),
\end{equation}
where $\mathbf{\mathsf{D}}_{ij} \equiv \partial_j \mathbf{f}_i \vert_{\bm{u}}$, yields
\begin{equation}
    \label{eq:loss_multivariate}
    \begin{split}
        \mathcal{L} \left( \bm{v} \right) &=
        \frac{1}{\sqrt{| 2 \pi \mathbf{\mathsf{\Sigma_\text{tot}}}|}}  \\
        &\times \exp \left(-\frac{1}{2} \left(\bm{v}-\bm{f}(\bm{u)}\right)^{\rm T} \mathbf{\mathsf{\Sigma}}_\text{tot}^{-1} \left(\bm{v}-\bm{f}(\bm{u)} \right) \right),
    \end{split}
\end{equation}
where $\mathbf{\mathsf{\Sigma_\text{tot}}}$ is the total covariance matrix defined by
\begin{equation}
    \label{eq:sigmatot_multivariate}
    \mathbf{\mathsf{\Sigma}}_\text{tot} \equiv
    \mathbf{\mathsf{\Sigma}}_v + \mathbf{\mathsf{D}} \mathbf{\mathsf{\Sigma}}_u \mathbf{\mathsf{D}}^{\rm T}.
\end{equation}
This is equivalent to Eq.~\ref{eq:sigmatot} if $\bm{u}$ and $\bm{v}$ have only a single element.
 
In principle, the assumption that the joint probability distributions of $\bm{u}^\text{t}$ and $\bm{v}^\text{t}$ are a multivariate Gaussian is unnecessary as the full empirical distributions are generated by the Monte Carlo sampling described above. Using this directly would enable a loss function which, while still independent between galaxies, fully encapsulates the correlated, non-Gaussian structure of each galaxy's measurements and hence provides a more accurate description of the expected probability distributions of $\gb$ and $\go$. This model would also allow for more accurate mock data generation as discussed in Sec.~\ref{sec:disc}.

\subsection{Sampling nuisance parameters from their posteriors}
\label{sec:app_post}

All the methods described above sample $\Upsilon$, $D$ and $i$, and the true $\gb$ values $\bm{u}^\text{t}$, from their priors, i.e. without adjusting them to maximise agreement with the function being fitted. In principle, the better procedure is to constrain these nuisance parameters jointly with any parameters of the function, assuming that function in the fit. Even the relatively simple case where $\bm{u}^\text{t}$ is drawn from the prior but $\Upsilon$, $D$ and $i$ from the posterior is however impractical as it requires optimisation in a parameter space of dimension $147\times3+p$ if the three $\Upsilon$s are coupled (or only one is varied), and $147\times5+p$ if they are varied separately, where $p$ is the number of free parameters in the function. This would naturally account for correlations between the $\gb$ and $\go$ measurements induced by variations in $\Upsilon$, $D$ and $i$, and thus provides a more accurate but expensive alternative to the model of Sec.~\ref{sec:app_corr}. This approach will be applied for the first time to the RAR, outside the context of SR, in upcoming work (Desmond 2023, in prep).


The situation is complicated further by inference of $\bm{u}^\text{t}$.
At the top level of the hierarchical model for predicting $\bm{v}$ from $\bm{u}$ are both the parameters of the model, $\bm{\theta}$, and $\bm{u}^\text{t}$. If one wanted to find the true maximum likelihood point, one should maximise Eq.~\ref{eq:uv_likelihood} for $\bm{u}^\text{t}$ as well as $\bm{\theta}$, instead of Eq.~\ref{eq:v_likelihood} for $\bm{\theta}$ alone.

To do this, we start by noting that maximising the likelihood in Eq.~\ref{eq:uv_likelihood} is equivalent to minimising
\begin{equation}
    \begin{split}
        h \left(\bm{u}^\text{t}, \bm{v} \right) &= 
        \frac{1}{2} \left( \bm{f} \left(\bm{u}^\text{t} \right) -\bm{v} \right)^{\rm T}  \mathbf{\mathsf{\Sigma}}^{-1}_v \left(\bm{f} \left(\bm{u}^\text{t} \right) -\bm{v} \right) \\
        & \quad + \frac{1}{2} \left(\bm{u}^\text{t}-\bm{u}\right)^{\rm T} \mathbf{\mathsf{\Sigma}}^{-1}_u \left(\bm{u}^\text{t}-\bm{u} \right).
    \end{split}
\end{equation}
Using Eq.~\ref{eq:taylor} and minimising $h$ with respect to $\bm{u}^\text{t}$, we find the maximum-likelihood point, $\hat{\bm{u}}^\text{t}$, to be
\begin{equation}
    \hat{\bm{u}}^\text{t}  = \bm{u}
    + \mathbf{\mathsf{\Sigma}}_u \mathbf{\mathsf{D}}^{\rm T}
    \mathbf{\mathsf{\Sigma}}_\text{tot}^{-1}
    \left( \bm{v} - \bm{f} \left(\bm{u} \right) \right),
\end{equation}
and we must therefore maximise
\begin{equation}
    \label{eq:mlp_multivariate}
    \begin{split}
        \mathcal{L} \left( \hat{\bm{u}}^\text{t}, \bm{v} \right) &=
         \frac{1}{\sqrt{ |2 \pi \mathbf{\mathsf{\Sigma}}_u|}}\frac{1}{\sqrt{| 2 \pi \mathbf{\mathsf{\Sigma}}_v|}}  \\
        &\times \exp \left(-\frac{1}{2} \left(\bm{v}-\bm{f}(\bm{u)}\right)^{\rm T} \mathbf{\mathsf{\Sigma}}_\text{tot}^{-1} \left(\bm{v}-\bm{f}(\bm{u}) \right) \right),
    \end{split}
\end{equation}
where $\mathbf{\mathsf{\Sigma}}_\text{tot}^{-1}$ is defined in Eq.~\ref{eq:sigmatot_multivariate}.
This will not yield the same result as maximising Eq.~\ref{eq:loss_multivariate} since Eq.~\ref{eq:mlp_multivariate} does not contain the normalisation term which penalises large gradients. When applied to mock data generated assuming $\bm{v}^\text{t}$ is linearly related to $\bm{u}^\text{t}$, we find that Eq.~\ref{eq:loss_multivariate} (the commonly used expression, and the one we adopt in Sec.~\ref{sec:loss}) induces a small bias that Eq.~\ref{eq:mlp_multivariate} does not. This is the reason why the best-fit RAR and generalised RAR IFs do not precisely match the generating function in Table~\ref{tab:mock_results}, or the Simple IF + EFE the generating function in Table~\ref{tab:mock_results_efe}. We should expect all best-fit parameter values to be likewise slightly biased by the use of Eq.~\ref{eq:loss}.

In practice, it is challenging to maximise Eq.~\ref{eq:uv_likelihood} since the lack of the gradient-penalising determinant means that na\"{i}vely using Eq.~\ref{eq:mlp_multivariate} can prefer functions with diverging gradients at at least one point in the domain of $\bm{u}$. This breaks the linearity assumption in Eq.~\ref{eq:taylor}, making the result untrustworthy. Instead, one should numerically solve the full optimisation problem without Taylor-expanding $\bm{f}$. This involves solving a root-finding problem for each trial $\bm{\theta}$ during the optimisation of the function's parameters. Although feasible for a handful of functions, this is computationally impractical for the full set of ESR functions. We therefore use the simpler likelihood here, calibrating our results using mock data to measure the magnitude of the bias. We defer further discussion of this important issue to future work.

\section{Limiting slopes of ESR Functions}
\label{sec:app_limits}

Here we provide the analytic low-$\gb$ ($s_-$) and high-$\gb$ ($s_+$) logarithmic slopes of the top 10 functions generated using ESR up to complexity 9. We give the slopes for the observed SPARC data in Table~\ref{tab:limits_data}, the mock data assuming the RAR IF in Table~\ref{tab:limits_rar}, and the mock data assuming the Simple IF with EFE in Table~\ref{tab:limits_efe}.

\begingroup
\setlength{\tabcolsep}{2pt}
\begin{table*}
    \label{tab:limits_data}
    \begin{center}
    \begin{tabular}{cc|cc|cc}
		\hline
		\multirow{2}{*}{Rank} & \multirow{2}{*}{Function} & \multicolumn{2}{c}{Low-acceleration slope} & \multicolumn{2}{c}{High-acceleration slope} \\
		& & Value & Condition & Value & Condition \\
		\hline
		1 & $\theta_0 \left(| \theta_1+x| ^{\theta_2}+x\right)$ & $0$ & & $\theta_2$ & $\theta_2 \geq 1$ \\
		& & & & $1$ & otherwise \\
		\hline
		2 & $\left| | \theta_1| ^x+\theta_0\right| ^{\theta_2}+x$ & $0$ & & $1$ & ($|\theta_1| \leq 1$) or ($|\theta_1| > 1$ and $\theta_2\leq0$) \\
		& & & & $\infty$ & otherwise \\
		\hline
		3 & $| \theta_0| ^{| \theta_1-x| ^{\theta_2}-\theta_3}$ & $0$ & & $\infty$ & $|\theta_0| > 1$ and $\theta_2 > 0$ \\
		& & & & $-\infty$ & $0 < |\theta_0| < 1$ and $\theta_2 > 0$ \\
		& & & & $0$ & otherwise \\
		\hline
		4 & $| \theta_0 (\theta_1+x)| ^{\theta_2}+x$ & $0$ & & $\theta_2$ & $\theta_2 \geq 1$ \\
		& & & & $1$ & otherwise \\
		\hline
		5 & $\left| \theta_0-| \theta_1-x| ^{\theta_2}\right| ^{\theta_3}$ & $0$ & & $\theta_2 \theta_3$ & $\theta_2 \geq 0$ \\
		& & & & $0$ & otherwise \\
		\hline
		6 & $\sqrt{x} \exp\left({\frac{| \theta_0+x| ^{\theta_1}}{2}}\right)$ & $\frac{1}{2}$ & & $\frac{1}{2}$ & $\theta_1 \leq 0$  \\
		& & & & $\infty$ & otherwise \\
		\hline
		7 & $\left(\frac{| \theta_0| ^x}{x}\right)^{\theta_1}+x$ & $-\theta_1$ & $\theta_1\geq-1$ & $1$ & ($|\theta_0|>1$ and $\theta_1\leq0$) or ($0<|\theta_0|<1$ and $\theta_1\geq0$) or ($|\theta_0|=1$ and $\theta_1\geq-1$)  \\
		& & $1$ & otherwise & $-\theta_1$ & $|\theta_0|=1$ and $\theta_1 < -1$ \\
		& & & & $1 + 0^{\theta_1} \theta_1 (-\infty)$ & $\theta_0=0$ \\
		& & & & $\infty$ & otherwise \\
		\hline
		8 & $\sqrt{| \theta_0+x| }+\theta_1 x$ & 0 & & $1$ & \\
		\hline
		9 & $\left| \theta_0+\frac{1}{\sqrt[4]{x}}\right| ^{\theta_1}$ & $-\frac{\theta_1}{4}$ & & $0$ & \\
		\hline
		10 & $\left(\sqrt{x}+\frac{1}{x}\right)^{\theta_0}+x$ & $-\theta_0$ & $\theta_0\geq -1$ & $\frac{\theta_0}{2}$ & $\theta_0 \geq 2$ \\
		& & $1$ & otherwise & $1$ & otherwise \\
		\hline
    \end{tabular}
	\caption{Functional forms and limiting slopes of the ten best functions found by ESR applied to the SPARC data. The functions in the second column give the fitted $y=\go / 10^{-10}$ m s$^{-2}$ for input $x=\gb/ 10^{-10}$ m s$^{-2}$. The low-acceleration slope is $\lim_{x\to0^+} \dd\log y / \dd \log x$ (denoted $s_-$ in the text), and the high-acceleration slope is similarly defined but for $x\to\infty$ ($s_+$). $\{ \theta_i \}$ are real parameters fitted to the data to maximise the likelihood (see Table~\ref{tab:real_results}). Slopes given without conditions are valid $\forall \,\theta_i$. For comparison, the MOND prediction without the external field effect is $s_-=1/2$ and $s_+=1$.}
	\end{center}
\end{table*}
\endgroup

\begin{table*}
    \label{tab:limits_rar}
    \begin{center}
    \begin{tabular}{*6c}
		\hline
		\multirow{2}{*}{Rank} & \multirow{2}{*}{Function} & \multicolumn{2}{c}{Low-acceleration slope} & \multicolumn{2}{c}{High-acceleration slope} \\
		& & Value & Condition & Value & Condition \\
		\hline
		1 & $\theta_0+\theta_1 x+\sqrt{x}$ & $0$ & & $1$ & \\
		\hline
		2 & $\sqrt{| \theta_0+x| }+\theta_1 x$ & $0$ & & $1$ & \\
		\hline
		3 & $\theta_0 x+x^{\theta_1}$ & $\theta_1$ & $\theta_1 \leq 1$ & $\theta_1$ & $\theta_1 \geq 1$\\
		& & $1$ & otherwise & $1$ & otherwise \\
		\hline
		4 & $\sqrt{x} \exp\left({\frac{x^{\theta_0}}{2}}\right)$ & $\frac{1}{2}$ & $\theta_0 \geq 0$ & $\frac{1}{2}$ & $\theta_0 \leq 0$ \\
		& & $-\infty$ & otherwise & $\infty$ & otherwise \\
		\hline
		5 & $(\theta_0+x) \left(\theta_1+\frac{1}{\sqrt{x}}\right)$ & $-\frac{1}{2}$ & & $1$ & \\
		\hline
		6 & $\frac{1}{\sqrt{\left| \theta_0+\frac{1}{x}\right| }}+x$ & $\frac{1}{2}$ & & $1$ & \\
		\hline
		7 & $(x | \theta_0| )^{(x | \theta_1| )^{\theta_2}}$  & $\infty$ & $|\theta_1| > 0$ and $\theta_2 < 0$ & $\infty$ & $|\theta_1| > 0$ and $\theta_2 > 0$ \\
		& & $1$ & $\theta_2=0$ & $0$ & $|\theta_1| > 0$ and $\theta_2 < 0$ \\
		& & $0$ & otherwise & $1$ & $|\theta_1| > 0$ and $\theta_2 = 0$ \\
		& & & & $0^{\theta_2} \theta_2 \infty$ & otherwise \\
		\hline
		8 & $\theta_0 x+| \theta_1+x| ^{\theta_2}$ & $0$ & & $\theta_2$ & $\theta_2 \geq 1$ \\
		& & & & $1$ & otherwise \\
		\hline
		9 & $x \left(| \theta_0-x| ^{\theta_1}-\theta_2\right)$ & $1$ & & $\theta_1 + 1$ & $\theta_1 \geq 0$ \\
		& & & & $1$ & otherwise \\
		\hline
		10 & $(\theta_0-x) \left(\theta_1-x^{\theta_2}\right)$ & $\theta_2$ & $\theta_2 \leq 0$ & $\theta_2 + 1$ & $\theta_2 \geq 0$  \\
		& & $0$ & otherwise & $1$ & otherwise \\
		\hline
    \end{tabular}
	\caption{As Table~\ref{tab:limits_data} but for the mock data generated from the RAR IF.}
	\end{center}
\end{table*}

\begingroup
\setlength{\tabcolsep}{2pt}
\begin{table*}
    \label{tab:limits_efe}
    \begin{center}
    \begin{tabular}{*6c}
		\hline
		\multirow{2}{*}{Rank} & \multirow{2}{*}{Function} & \multicolumn{2}{c}{Low-acceleration slope} & \multicolumn{2}{c}{High-acceleration slope} \\
		& & Value & Condition & Value & Condition \\
		\hline
		1 & $\theta_0+\sqrt{x^2+2 x}$ & $0$ & & $1$ & \\
		\hline
		2 & $\theta_0+\sqrt{x | \theta_1+x| }$ & $0$ & & $1$ & \\
		\hline
		3 & $-| \theta_0| ^{\sqrt{x}}+\theta_1+x$ & $0$ & & $1$ & $|\theta_0| \leq 1$ \\
		& & & & $\infty$ & otherwise \\
		\hline
		4 & $(\theta_0-x) \left(\theta_1-x^{\theta_2}\right)$ & $\theta_2$ & $\theta_2 \leq 0$ & $1+\theta_2$ & $ \theta_2\geq0$ \\
		& & $0$ & otherwise & $1$ & otherwise \\
		\hline
		5 & $x^{\theta_0}-\theta_1 (\theta_2-x)$ & $\theta_0$ & $\theta_0 \leq 0$ & $\theta_0$ & $\theta_0 \geq 1$ \\
		& & $0$ & otherwise & $1$ & otherwise \\
		\hline
		6 & $| \theta_0-x| ^{\theta_1}-\theta_2 x$ & $0$ & & $\theta_1$ & $\theta_1 \geq 1$ \\
		& & & & $1$ & otherwise \\
		\hline
		7 & $x | \theta_0| ^{-| \theta_1| ^{x^{\theta_2}}}$ & $\infty \log|\theta_0|$ & $|\theta_1| > 1$ and $\theta_2 < 0$ & $-\infty$ & $|\theta_0|>1$ and $|\theta_1|>1$ and $\theta_2>0$  \\
		& & $1$ & otherwise & $\infty$ & $0<|\theta_0|<1$ and $|\theta_1|>1$ and $\theta_2>0$ \\
		& & & & $\infty \theta_2$ & $\theta_0=0$ and $|\theta_1| > 1$ \\
		& & & & $1$ & otherwise \\
		\hline
		8 & $x \left(\theta_0+| \theta_1+x| ^{\theta_2}\right)$ & $1$ & & $1+\theta_2$ & $\theta_2 \geq 0$ \\
		& & & & $1$ & otherwise \\
		\hline
		9 & $| \theta_0| ^{| \theta_1| ^{x^{\theta_2}}}+x$ & $-\infty$ & $|\theta_0|>1$ and $|\theta_1|>1$ and $\theta_2<0$ & $\infty$ & $|\theta_0|>1$ and $|\theta_1|>1$ and $\theta_2>0$ \\
		& & $1$ & ($\theta_0=0$) or ($\theta_1=0$) or ($|\theta_0|<1$ and $|\theta_1|>1$ and $\theta_2<0$)& $1$ & otherwise \\
		& & $0$ & otherwise & & \\
		\hline
		10 & $\exp\left({\theta_0-\frac{1}{\sqrt[4]{x}}}\right)+x$ & $1$ & & $1$ & \\
		\hline
    \end{tabular}
	\caption{As Table~\ref{tab:limits_data} but for the mock data generated using the Simple IF + EFE.}
	\end{center}
\end{table*}
\endgroup

\bsp
\label{lastpage}
\end{document}